\documentclass[]{aa}

\usepackage{graphicx}
\usepackage[colorlinks,allcolors=blue]{hyperref}
\usepackage{xurl}
\usepackage{txfonts}
\usepackage[rightcaption]{sidecap} 

\begin{document}

   \title{Investigating the link between inner gravitational potential and star-formation quenching in CALIFA galaxies}
   \subtitle{}
   \author{V.  Kalinova\inst{1}
           \and D. Colombo\inst{1}
           \and S. F. S{\'a}nchez\inst{2}
           \and E. Rosolowsky\inst{3}
           \and K. Kodaira\inst{1,4,5}
           \and R. Garc{\'i}a-Benito\inst{6}, \\
                S. E. Meidt\inst{7}
        \and    T. A. Davis\inst{8}
          \and  A. B. Romeo\inst{9}
          \and S.-Y. Yu\inst{1}
          \and  R. Gonz{\'a}lez Delgado\inst{6}
           \and E. A. D. Lacerda\inst{2}
          }

   \institute{Max Planck Institute for Radioastronomy, Auf dem H\"ugel 69, D-53121 Bonn, Germany \\
   email: \texttt{kalinova@mpifr.de}
        \and Instituto de Astronom\'\i a,Universidad Nacional Auton\'oma de M\'exico,   A.P. 70-264, 04510, M\'exico, D.F.
         \and Department of Physics 4-181 CCIS, University of Alberta, Edmonton AB T6G 2E1, Canada 
         \and National Astronomical Observatory of Japan;  Osawa2-21-1, Mitaka-shi, Tokyo, Japan  PC 181-8588
         \and SOKENDAI, International Village, Hayama-machi, Miura-gun, Kanagawa-ken, Japan, PC 240-0193
         \and Instituto de Astrof\'isica de Andaluc\'ia, CSIC, Apartado de correos 3004, 18080 Granada, Spain 
         \and Sterrenkundig  Observatorium,  Universiteit  Gent,  Krijgslaan  281  S9, B-9000 Gent, Belgium
         \and School of Physics and Astronomy, Cardiff University, Queens Buildings, The Parade, Cardiff CF24 3AA, UK
         \and Department of Space, Earth and Environment, Chalmers University of Technology, SE-41296 Gothenburg, Sweden
             }

   \date{Received March 14, 2022; accepted July 7, 2022}

 
\abstract{It has been suggested that the gravitational potential can have a significant role in suppressing the star formation in the nearby galaxies. To establish observational constrains on this scenario, we investigate the connection between the dynamics, through the circular velocity curves (CVCs) as a proxy of the inner gravitational potential, and star formation quenching in 215 non-active galaxies across Hubble sequence from the Calar Alto Legacy Integral Field Area (CALIFA) survey. 
Our results show that galaxies with similar CVCs tend to have a certain star-formation quenching pattern. 
To explore these findings in more details, we construct kpc-resolved relations of the equivalent width of the H$\alpha$ ($W_{{\rm H}\alpha}$) versus the amplitude  ($V_c$) and shape ($\beta= d\ln V_c/ d\ln R$)  of the circular velocity at given radius.
We find that the $W_{{\rm H}\alpha}-V_c$ is a declining relationship, where the retired regions of the galaxies (the ones with $W_{{\rm H}\alpha}$ values below 3 {\AA}) tend to have higher $V_c$.
Differently, $W_{{\rm H}\alpha}-\beta$ is a bi-modal relationship, characterised by two peaks: concentration of the star forming regions at a positive $\beta$ (rising CVC) and another one of the retired regions with a negative $\beta$ (declining CVC). 
Our results show that both the amplitude of the CVC, driven by the mass of the galaxies, and the shape of the CVC, reflecting the internal structure of the galaxies, play an important role in galaxy's quenching history.}


   \maketitle
%


\defcitealias{Kalinova2017b}{K17}
\defcitealias{Kalinova2021}{K21}


\section{Introduction}
\label{S:intro}
The transition of the galaxies from blue and young to red and retired systems is referred to the process "star-formation quenching" (e.g. \citealt{Strateva2001, Faber2007}), which is  characterised by a fast decline in the star formation rate (SFR; e.g. \citealt{Corcho-Caballero2021}). 
Several quenching mechanisms have been proposed to explain the diversity of galaxies we observe today and they can be grouped to \emph{external} (outside of the galaxy) and \emph{internal} (within the galaxy).\\
\emph{Environmental quenching} (\citealt{Peng2010},\citealt{Gunn1972, Abadi1999}) is an external process. This includes ram-pressure stripping of the galaxies interstellar medium, strangulation (removal of the outer gaseous envelope of the galaxy falling into galaxy cluster, which will cease the  star formation of the galaxy as the only available gas is hot and it cannot replenish its cold gas reservoir (\citealt{Larson1980, Balogh2000}) and galaxy harassment (quenching due to interactions with other members of the galaxy cluster, leading to dynamical heating \citealt{Farouki1981,Moore1996, Bluck2020b,Bluck2020a}). Low-mass galaxies, in particular, can have their gas removed by tidal stripping during travel towards the centre of a galaxy cluster \citep[e.g. ][]{Abadi1999}.

In addition, a large range of internal processes are also thought to be able to quench star formation.
\emph{Dark matter (DM) halo quenching} is considered an internal process, related to the most extended component of the galaxies.  In systems with critical DM halo mass of $\sim$10$^{12}$ M$_{\odot}$, the accreated gas is shocked and heated to the virial temperature of the halo, preventing star formation (\citealt{Birnboim2003}). The correlation of the stellar component with the quenching of the massive galaxies is known as \emph{mass quenching} (\citealt{Peng2010}).  
\emph{Active galactic nuclei (AGN)} feedback quenching is able to suppresses star formation in the galaxies (\citealt{Husemann2018}). AGN feedback can transfer radiation to the surrounding gas and suppress gas accretion \citep{DiMatteo2005} or kinetic energy and momentum can cause expulsion of gas \citep{Croton2006}. The ejective galactic winds from supernovae or HII regions constitute another quenching feedback mechanism (\emph{stellar feedback quenching}) that acts mostly in low-mass and late-type galaxies (e.g. \citealt{Colling2018}). 
Internal dynamics contributes to the quenching of the galaxies too.
Secular evolution processes can form a bar structure that generates radial inflow towards the centre of the galaxies, increasing the random motion of the gas and the disc heating that stabilises the inner gaseous disc \citep{Romeo2015,Romeo2016,Khoperskov2018}. Additionally bars drive gas flow toward galactic centre to trigger a central star-burst, which in turn can consume almost all the gas causing a quenched centre. This action leads to a periodically episodes of quenching and star formation at the centre of bars \citep{Krumholz2015}.
The growth of the central spheroid (bulge) seems to play an important role in ceasing the star formation in the galaxies via stabilisation of the disk against gravitational instability, known as \emph{"morphological quenching"} (\citealt{Martig2009}), gravitational quenching (\citealt{Genzel2014}) or \emph{dynamical suppression} (e.g. \citealt{Davis2014,Gensior2020,Gensior2021}). 

\begin{figure}
\includegraphics[width=0.5\textwidth]{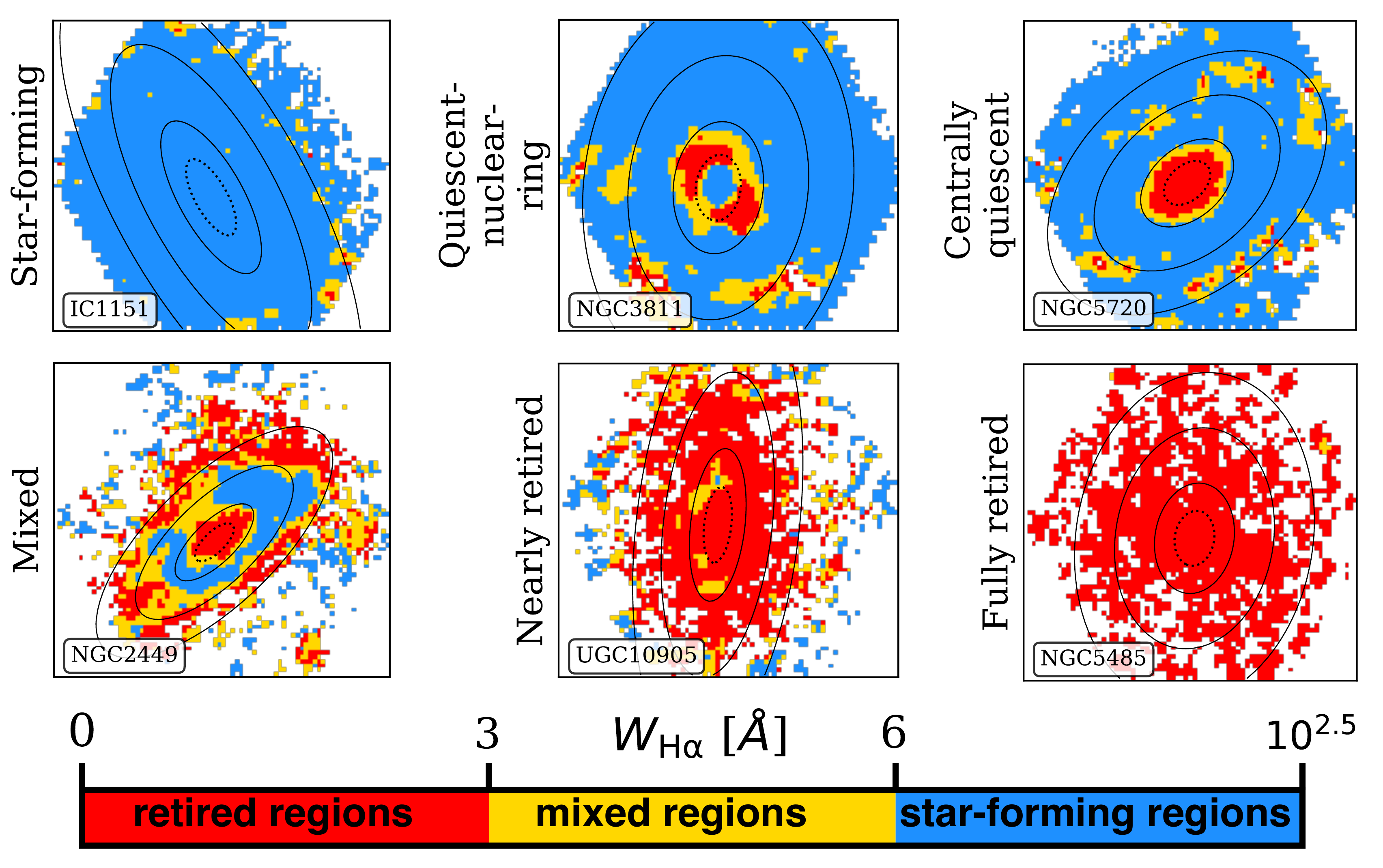}
\caption{Example of galaxy quenching stages defined by \citetalias{Kalinova2021} through visual inspection of the W$_\mathrm{H\alpha}$ distribution (from top left to bottom right): \emph{star-forming}, \emph{quiescent-nuclear-ring}, \emph{centrally quiescent}, \emph{mixed}, \emph{nearly retired}  and \emph{fully retired}. The colour bar separates the regions of the W$_\mathrm{H\alpha}$ maps (e.g. \citealt{Sanchez2014}, \citetalias{Kalinova2021}) on: star-forming (W$_\mathrm{H\alpha}>$ 6 {\AA}), mixed ($3 <\mathrm{W}_\mathrm{H\alpha} \leq$ 6 {\AA}) and retired (W$_\mathrm{H\alpha} \leq$ 3 {\AA}). The dashed contours correspond to 0.5 of the galaxy's effective radius (R$_\mathrm{e}$), while the continuous contours indicate 1, 2 and 3R$_\mathrm{e}$. See Sect. \ref{S:intro} and \citetalias{Kalinova2021} for more details.}
\label{fig:QSC}
\end{figure}

Overall, the causes of star-formation quenching in galaxies are  highly complex and still debated. It is not clear whether there is one dominant mechanism or a mixture of several processes that are responsible for the variety of galaxy morphologies seen today.
Recent resolved studies have tried to disentangle these effects by classifying the galaxies based on their star formation activity (or its absence; \citealt{Singh2013,Belfiore2016,Lacerda2018}). In particular, \cite{Kalinova2021} (hereafter \citetalias{Kalinova2021}) distinguish various quenching patterns (called "quenching stages"; Fig. \ref{fig:QSC}) in the spatially resolved ionised gas distribution of the galaxies, where the ratio star-forming versus quenching regions of the galaxies increases from late-type to early-type galaxy morphologies (e.g. \citealt{Lacerda2018}, \citetalias{Kalinova2021}).
Based on H$\alpha$-equivalent-width  ($W_{{\rm H}\alpha}$) thresholds (e.g., \citealt{Sanchez2014}, \citealt{Lacerda2020}), \citetalias{Kalinova2021} distinguished star-forming  ($W_{\rm{H_{\alpha}}} > $ 6 {\AA}), mixed  (3 $<$ $W_{\rm{H_{\alpha}}} \leq $ 6 {\AA}) and retired ($W_{\rm{H_{\alpha}}} \leq $ 3 {\AA}) regions  within the field of view of the galaxies. They proposed six quenching stages of the systems (see Fig. \ref{fig:QSC}): \emph{star-forming} (fully dominated by recent star-formation), \emph{quiescent-nuclear-ring} (presence of a quiescent-ring structure in the central regions, but still dominated by star formation in the outskirts), \emph{centrally quiescent} (quiescent inner region within $0.5 R_e$ of the galaxy, where $R_e$ is the effective radius of the galaxy), \emph{mixed} (no clear patterns in the ionised gas distributions), \emph{nearly retired} (quiescent galaxies with little star-formation regions) and \emph{fully retired} (completely quiescent objects up to $2R_e$). 

In this study, we explore the connection between the dynamics and  star-formation quenching stage of the galaxies, shedding light on the role of the dynamical suppression mechanism for their formation. To achieve this goal, we compare the  circular velocities of the galaxies (tracing the total gravitational potential of the systems; \citealt{Kalinova2017b}, hereafter \citetalias{Kalinova2017b}) and the values of the W$_\mathrm{H\alpha}$ (\citetalias{Kalinova2021}), serving as a star-formation/quenching marker across Hubble sequence.

\sidecaptionvpos{figure}{c}

\section{Sample and Data}
\label{S:data}
Our sample is derived from the 238 CALIFA (Calar Alto Legacy Integral Field Area; \citealt{Sanchez2012}) survey galaxies, extensively analysed in  \citetalias{Kalinova2017b} and \citetalias{Kalinova2021}. It  is a representative population of the CALIFA mother sample (see Fig. 1 in \citetalias{Kalinova2017b}), and therefore of the nearby Universe galaxies (\citealt{Walcher2014}).
In our analysis, we only discard few targets - the active galaxies members (15 strong AGN and 8 weak AGN galaxies) from the original sample of \citetalias{Kalinova2017b} and \citetalias{Kalinova2021} to avoid any possible biases due to the presence of nuclear activity in the galaxies. The final sample consists of 215 non-active galaxies, spanning over six different quenching stages as defined in \citetalias{Kalinova2021} (see Fig. 1 of \citetalias{Kalinova2021} and Fig. \ref{fig:QSC} of this study), various morphologies (from elliptical to late-type spiral galaxies), stellar masses (from 6$\times$$10^8$ $M_{\odot}$ to 5$\times$$10^{11}$ $M_{\odot}$) and redshifts ($0.005<z<0.03$). 
Further details about the survey and data reduction can be found in \cite{Sanchez2012}, \cite{Husemann2013}, \cite{Garcia-Benito2015} and \cite{Sanchez2016c}.

To perform this study, we use the publicly available circular velocity curve (CVC) catalogue of \citetalias{Kalinova2017b}. CVC is defined as $V_c^2 (R)=R\frac{\partial \Phi}{\partial R}{\vert_{z=0}}$, where $\Phi(R,z)$ is the gravitational potential and $R$ is the galactocentric radius, respectively.
Further, the CVC is calculated through the solutions of the axisymmetric Jeans equations by applying the Jeans Axisymmetric Modelling (JAM) code of \cite{Cappellari2008}\footnote{http://purl.org/cappellari/software}. 
First, \citetalias{Kalinova2017b} derive the surface brightness (SB) of the galaxies using the {\it r}-band images from Sloan Digital Sky Survey (SDSS)\footnote{https://www.sdss.org/}  catalogue of Data Release 12 (DR12; \citealt{Alam2015}) through the multi-Gaussian expansion method (MGE; \citealt{Monnet1992,Emsellem1994}). Given the defined SB of the galaxies,  the second velocity moment of the galaxies ($V_{\rm rms}=\sqrt{V^2+\sigma^2}$) of the stellar kinematics (based on the high-resolution ``V1200'' dataset; \citealt{Falcon-Barroso2017}) is fitted,  assuming constant dynamical mass-to-light ratio ($\gamma_{\rm{dyn}}$) and constant velocity anisotropy ($\beta_z=1-\sigma_z^2/\sigma_r^2$)  in the galactic  meridional plane. 
In the end, the CVCs were inferred from the de-projected SB of the galaxies, scaling by the best fit value of $\gamma_{\rm{dyn}}$ (see eq. 3 in \citetalias{Kalinova2017b}).

The emission-line analysis for deriving
the maps of the equivalent width of H$\alpha$ line ($W_{\rm{H\alpha}}$) was based on the {\sc Pipe3D} pipeline (\citealt{Sanchez2016a,Sanchez2016b}) calculations, assuming a Salpeter (\citealt{Salpeter1955}) initial mass function (IMF). Further details about the adopted data and analysis are provided in \citetalias{Kalinova2017b} and \citetalias{Kalinova2021}.

\section{Results}
\label{S:results}
\begin{figure*}
\centering
\includegraphics[width=1\textwidth]{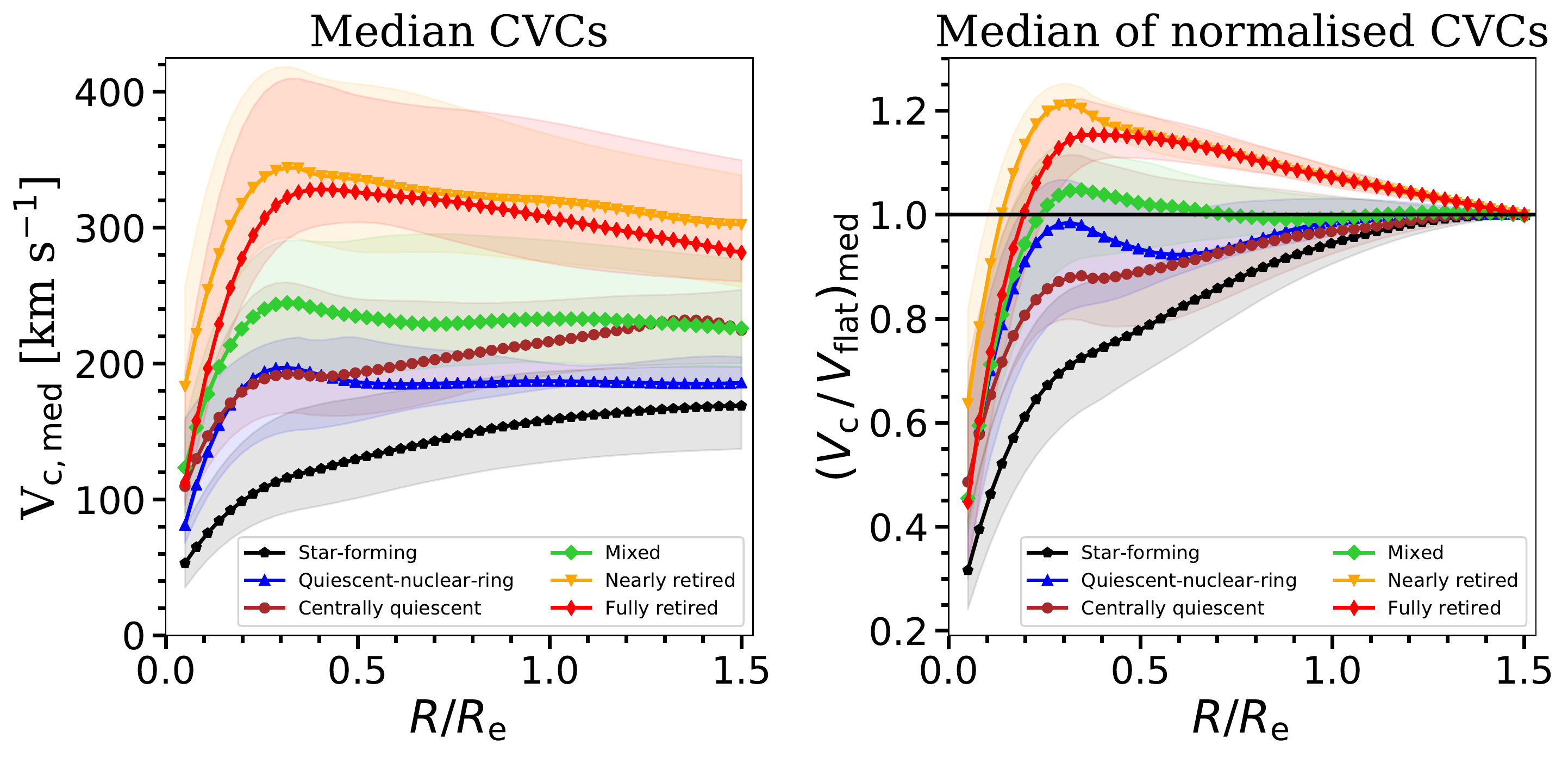}
\caption{
\emph{Left:} Median curve of the CVC profiles for each quenching stage group from star-forming to fully retired galaxies. \emph{Right:} Median of the CVC profiles, normalised with respect to the asymptotic velocity for each quenching stage group. The uncertainty bands correspond to the 25th percentile (below) and the 75th percentile (above) of the median distribution. To smooth the median CVCs and the  percentile profiles, we apply the Savitzky-Golay smoothing filter (\citealt{Savitzky-Golay1964}) by adopting third degree polynomial and 21-point wide sliding window (see Sect. \ref{S:results}).}
\label{fig:Vc_prototype}
\end{figure*}
\begin{figure*}
\centering
\includegraphics[width=1\textwidth]{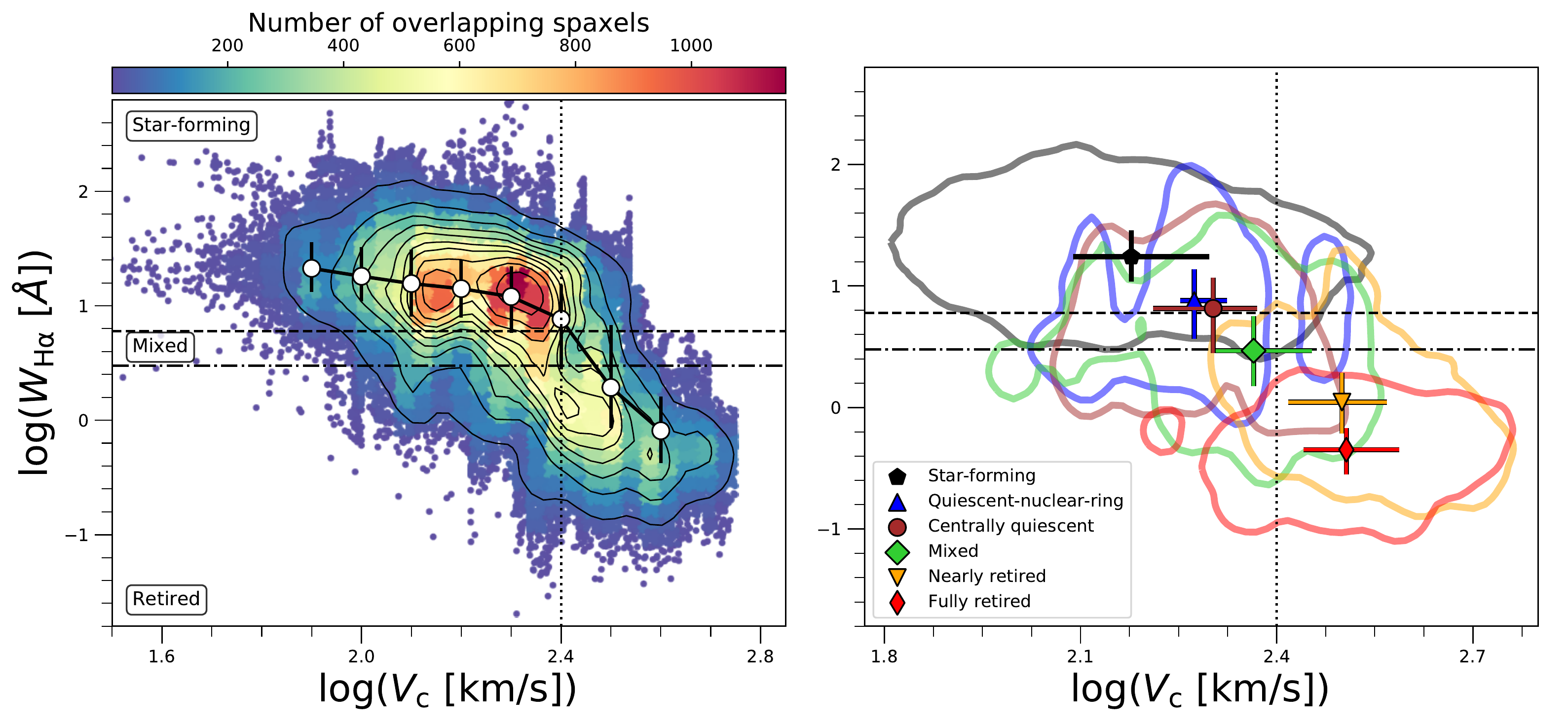}
\caption{CVC amplitude versus the W$_{\alpha}$ value, which defines the quenching, mixed and star-forming regions of the galaxies at each position of the sky.  
\emph{Left panel:} bi-dimensional histograms (colours) overlaid with black contours indicating the regions that contains between 15\% to 90\% of the total amount of points. White circles indicate the sample $W_{\rm H\alpha}$ medians within bins of 0.1 dex, while the vertical black bars encompass the 25$^{\rm th}$ to the 75$^{\rm th}$ percentiles of the distributions within the bins. 
(Note that the estimated median values do not generally correspond to the peaks of the spaxel distribution, indicated by the redder colours or by the contour levels, due to non-Gaussian distributions of the $W_{\rm H\alpha}$ values within given $V_c$ bins).
The colour bar indicates the number of the overlapping spaxels. \emph{Right panel:} Coloured contours contain 95\% of the data points within a given quenching stage. With the same colours, the markers indicates the position of the medians of $W_{\rm H\alpha}$ and $V_{\rm c}$ distributions of each stage, while the coloured bars the extension of the distributions within the 25$^{\rm th}$ and 75$^{\rm th}$ percentiles. Dashed-dotted and dashed black horizontal lines indicate the positions of $W_{\rm H\alpha} = 3\,\AA$ and $W_{\rm H\alpha} = 6\,\AA$, respectively. The vertical dotted line marks the critical velocity of $\log$($V_c$)$\sim$2.4 km s$^{-1}$ (V$_c\sim$ 250 km s$^{-1}$) at which the relation start to steepen. }
\label{fig:Vc_Wha}
\end{figure*}
\begin{figure*}
\centering
\includegraphics[width=1\textwidth]{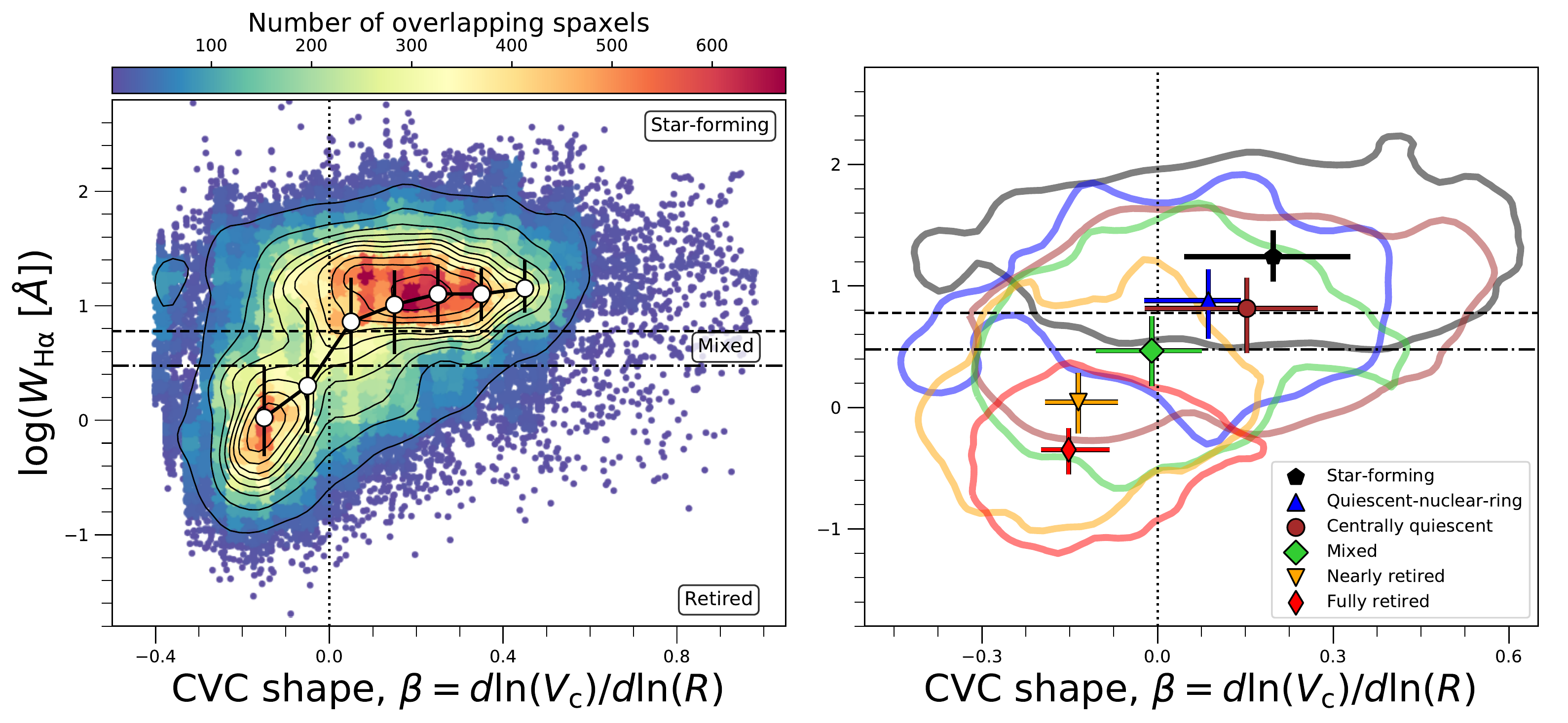}
\caption{Quenching versus shape of the CVC measured through $\beta=d\ln V_c/d\ln R$, which value we infer for each pixel of the radial and circular velocity maps. The rising part of the curve behaves as a solid body, $V_c\propto R$, giving $\beta=1$, while in the flat part of the curve, $V_c=$constant, and $\beta=0$ (vertical dotted line). Negative value of the logarithmic derivative ($\beta<0$) corresponds to the declining part of the CVC. Symbols and conventions follow Fig.~\ref{fig:Vc_Wha}.}
\label{fig:Beta_Wha}
\end{figure*}
\begin{figure*}
\centering
\includegraphics[width=0.96\textwidth]{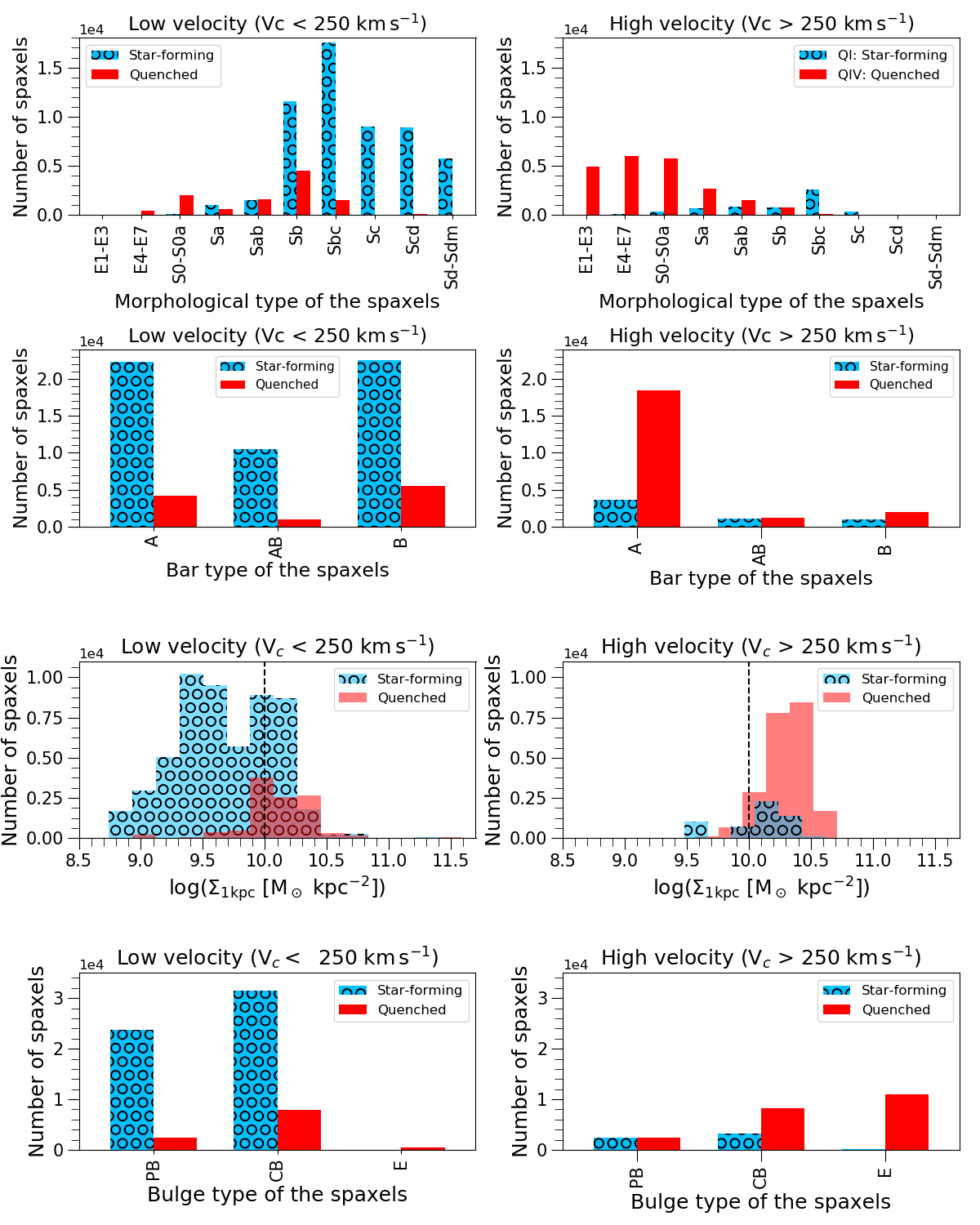}
\caption{Spaxel distribution of $W_{\rm H\alpha}-V_c$ relation across properties of the host galaxies. The first and the second row panels correspond to the morphological and bar-type (A: no bar, B: bar, and AB: unsure bar) classifications of the galaxies from \cite{Walcher2014}, respectively. The third row panels explore the central stellar mass surface density (\citealt{Sanchez2016a,Sanchez2016b}) within 1 kiloparsec ($\Sigma_{\mathrm{1kpc}}$) of the host galaxies, following \cite{Cheung2012}. The fourth row panels relates to the bulge type of the host galaxies, according to the method described in \citet[][PB: pseudo bulge, CB: classical bulge, and E: ellipticals]{Luo2020}. The left column panels explore the statistics of the spaxels for low velocities (V$_c <$ 250 km s$^{-1}$), while the right panels correspond to the high velocities (V$_c >$ 250 km s$^{-1}$). The different colours and symbols of the bars distinguish star-forming (blue with black circle) and quenched (red with no symbol) regions of the galaxies. The dashed horizontal lines at $\Sigma_{\mathrm{1kpc}}$=10$^{10}$ M$_\odot$ kpc$^{-2}$ are drawn to guide the eye. See more details in Sec. \ref{SS:scatter}.} 
\label{fig:Vc_stat}
\end{figure*}


\begin{figure*}
\centering
\includegraphics[width=0.96\textwidth]{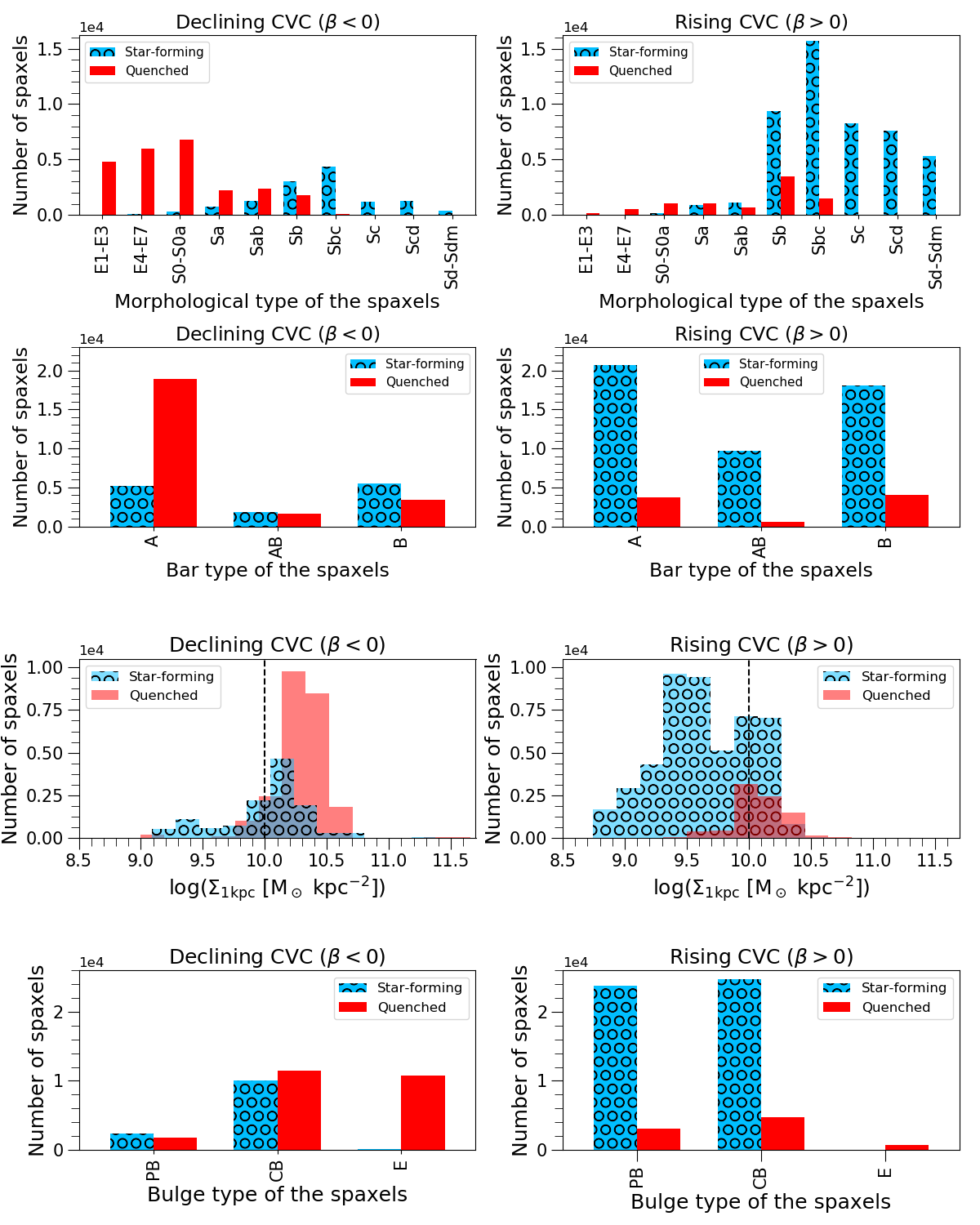}
\caption{Spaxel distribution of $W_{\rm H\alpha}-\beta$ relation across properties of the host galaxies. The first and the second row panels correspond to the morphological and bar-type (A: no bar, B: bar, and AB: unsure bar) classifications of the galaxies from \cite{Walcher2014}, respectively. The third row panels explore the central stellar mass surface density (\citealt{Sanchez2016a,Sanchez2016b}) within 1 kiloparsec ($\Sigma_{\mathrm{1kpc}}$) of the host galaxies, following \cite{Cheung2012}. The fourth row panels relates to the bulge type of the host galaxies, according to the method described in \citet[][PB: pseudo bulge, CB: classical bulge, and E: ellipticals]{Luo2020}.  The left column panels explore the statistics of the spaxels for the declining CVCs ($\beta < $ 0), while the right panels correspond to the rising CVCs ($\beta >$ 0). 
The different colours and symbols of the bars in the panels distinguish star-forming (blue with black circle) and quenched (red with no symbol) regions of the galaxies. The dashed horizontal lines at $\Sigma_{\mathrm{1kpc}}$=10$^{10}$ M$_\odot$ kpc$^{-2}$ are drawn to guide the eye. See more details in Sec. \ref{SS:scatter}. }
\label{fig:beta_stat}
\end{figure*}

Our first  test towards understanding the link between the inner gravitational potential of the galaxies and their quenching stage is performed in Fig. \ref{fig:Vc_prototype}.  
We group the CVCs of the galaxies by their quenching stage using the CALIFA CVCs catalogue of \citetalias{Kalinova2017b}.
The amplitude and the slope of the central regions ($\sim$ 0.1--0.4 $R_e$) of the median profiles (left panel) increase as the galaxies become progressively more quenched, e.g., increasing in the sequence {\it star-forming} to {\it quiescent-nuclear-ring}/{\it centrally quiescent} to {\it mixed} to {\it fully retired} to {\it nearly retired}. 
We notice that in the central region (up to 0.4 $R_e$), centrally quiescent systems have similar median velocity as the quiescent-nuclear-ring galaxies, but larger velocity in the outer parts (above 0.4 $R_e$). This indicates that the disc of the centrally quiescent galaxies is more massive than the one of the quiescent-nuclear-ring systems.
Contrary, in the outer regions (above 0.4$-$0.5 $R_e$), the nearly retired galaxies have similar disc median velocity as the fully retired systems, but larger bulge median velocity in the central parts (below 0.4$-$0.5 $R_e$).
Among all quenching-stage groups, the nearly retired galaxies have the largest central mass concentration (the highest $V_c-$peak).
This might suggest that the fully retired systems (or even the rest of the quenching stage galaxies) are  progenitors of the nearly retired galaxies (via re-ignition of star-formation and mass build-up by merger events), or the nearly retired class has followed a different evolutionary path from the rest of the classes.
Furthermore, in the right panel of Fig. \ref{fig:Vc_prototype}, the 
median of the CVC profiles, normalised with respect to the asymptotic velocity, show more distinguishable order: {\it star-forming} to {\it centrally quiescent} to {\it quiescent-nuclear-ring} to {\it mixed} to {\it fully retired} to {\it nearly retired}.
Nevertheless, individually, the curves in a given quenching stage show a large variety. In particular, several star-forming galaxies are characterised by declining profiles, while some retired objects have rising profiles (see also the individual CVCs in each quenching-stage group in Figs. \ref{fig:Vc_EL} and \ref{fig:Vc_EL_flat}).

Figure \ref{fig:Vc_prototype} suggests that on average, there is a correlation between the CVC shape/amplitude and the quenching stage. To have a more detailed quantification of these correlations, we construct two resolved relationships using the available information from the $W_{\rm{H_{\alpha}}}$ maps and CVCs.
\subsection{Resolved quenching relations}
\label{SS:resolved}
The first relation, quenching-velocity ($W_{\rm{H_{\alpha}}}-V_c$), compares the  $W_{\rm{H_{\alpha}}}$ value 
and the amplitude of the circular velocity ($V_{\rm c}$) at a given position in the field of view of the galaxy (i.e. spaxel-by-spaxel). We calculate the circular velocity, $V_{\rm c}$, value at each radius of the de-projected radial map of the galaxies using the \texttt{MGE\_VCIRC} procedure included in the JAM package of \cite{Cappellari2008}. The inclination and the MGE models of the galaxies were adopted from  \citetalias{Kalinova2017b}. Assuming that the galaxies are axisymmetric, we derive ``maps'' with constant values of $V_{\rm c}$  at a given galactocentric radius. 
Once we obtain the $V_{\rm c}$ map, we are able to construct the second relationship, quenching$-$CVC shape ($W_{\rm{H_{\alpha}}}-\beta$), where $\beta$ is the derivative of $V_{\rm c}$ with respect to the radius, 
$\beta= d\ln V_c/ d\ln R$.
If the rising part of the CVC is close to a solid body rotation, then $V_c\propto R$, giving $\beta=1$, while if the part of the curve is flat gives $V_c=$constant, and therefore $\beta=0$.
If the logarithmic derivative of the CVC is negative ($\beta$<0) then the CVC is declining (e.g. Sect. 4.2.4 of \citealt{Leroy2008}). The statistical robustness of both resolved relationships is demonstrated in Appendix~\ref{A:realisations}.

Additionally, we use the Pearson’s correlation coefficient\footnote{\url{https://docs.scipy.org/doc/scipy/reference/generated/scipy.stats.pearsonr.html}} ($\rho$; e.g.  \citealt{Kowalski1972}) to test the strength of the monotonically linear relation. The coefficient $\rho$ can vary between +1 and -1 for linear correlation and linear anti-correlation, respectively (where 0 refers to no correlation between the two studied data sets). The Spearman’s rank correlation coefficient\footnote{\url{https://docs.scipy.org/doc/scipy/reference/generated/scipy.stats.spearmanr.html}} ($s$; e.g. \citealt{Zwillinger1999}) tells us whether the tested relationship is monotonically increasing ($s=1$), monotonically decreasing ($s=-1$) or there is non-monotonic relationship ($s=0$), independently from its linearity.

The \emph{resolved quenching-velocity relation}, $W_{\rm{H_{\alpha}}}-V_c$, which includes all values of the spaxels in each galaxy of our sample is displayed in Fig. \ref{fig:Vc_Wha} (left panel). 
Generally, the two quantities are moderately correlated
($\rho$=$-$0.62 and $s$=$-$0.60 ) with high significance (for both coefficients, we obtained p-value $\ll0.01$ indicating that the moderate correlation between the two quantities is not due to chance). The relation presents two slopes, as it flattens in low velocities and steepens in the high-velocities above the critical value $\log$($V_c$)$\sim$2.4 km s$^{-1}$ ($V_c\sim$ 250 km s$^{-1}$). 
\emph{Overall, the retired spaxels (i.e. those with $W_{\rm{H_{\alpha}}} < $ 3 {\AA}) tend to be characterised by higher circular velocities.} 
Furthermore, if we segregate the quenching-velocity relation into the six quenching stages (right panel of Fig. \ref{fig:Vc_Wha}; see also Fig. \ref{fig:wha_vc_qs}), we observe a smooth transition from star-forming to fully retired galaxies from the top-left (high $W_{\rm{H_{\alpha}}}$ and low $V_c$) to the bottom-right (low $W_{\rm{H_{\alpha}}}$ and high $V_c$) side of the relation. The central regions of the relation ($\log(W_{\rm{H_{\alpha}}}) \sim $ 1 {\AA} and $\log(V_c) \sim$ 2.3 km s$^{-1}$) are occupied by the mixed group of galaxies, which are actually green valley galaxies on the way to be fully quenched (see Fig. 11 of \citetalias{Kalinova2021}).

In Fig. \ref{fig:Beta_Wha}, the $W_{\rm{H_{\alpha}}}$ and $\beta$ values moderately correlate ($\rho$=0.49 and $s$=0.48),  as for the previous relation, with high significance (p-value $\ll0.01$ for both tests). Contrary to the $W_{\rm{H_{\alpha}}}-V_c$ relation, the $W_{\rm{H_{\alpha}}} - \beta$ relation is bi-modal and characterised by two clear peaks, where one of them is located at $\beta \sim -0.25$ and $W_{\rm{H_{\alpha}}} \sim 1 \AA$, and the second one is located at $\beta \sim 0.25$ and $W_{\rm{H_{\alpha}}}>6\,\AA$, indicating that generally star forming regions are characterised by rising rotation curves, while retired regions by declining profiles.

We also divide the $W_{\rm{H_{\alpha}}}-\beta$ diagram across the six quenching stages (see right panel of Fig. \ref{fig:Beta_Wha} and Fig. \ref{fig:wha_beta_qs}).  There is a smooth transition from fully retired to star-forming galaxies from the left-bottom (low $W_{\rm{H_{\alpha}}}$ and negative $\beta$) to the right-top side (high $W_{\rm{H_{\alpha}}}$ and positive $\beta$) of the relation. The intermediate quenching stage spaxel distribution  of the mixed galaxies is centrally located at the $W_{\rm{H_{\alpha}}}-\beta$  plane due to the large fraction of flat CVCs that these systems possess (see Fig. \ref{fig:Vc_prototype}, Fig. \ref{fig:Vc_EL} and Fig. \ref{fig:Vc_EL}).  In particular, \emph{the nearly retired and fully retired quenching stage galaxies are almost fully located in the quenched region with  declining CVC of the $W_{\rm{H_{\alpha}}}-\beta$ diagram (i.e. $W_{\rm{H_{\alpha}}}$ < 3 {\AA}  and $\beta<0$).}

\subsection{Nature of the scatter in the quenching relations}
\label{SS:scatter}
In the previous section, the $W_{\rm H\alpha}-V_c$ relation depicts a link between quenching and the amplitude of CVCs, showing in particular that a majority of spaxels with high velocity (i.e. $V_c\sim$ 250 km s$^{-1}$) are quenched (i.e.  $W_{\rm H\alpha} <$ 3 {\AA} ). The $W_{\rm H\alpha}-\beta$ relation, meanwhile, highlights the connection between quenching and the shapes of CVCs and exhibits a clear bi-modality: whereas star forming regions are overall characterised by rising curves, quenched regions are associated with decreasing curves.

Both quenching relations exhibit a large level of scatter, however, in the $W_{\rm H\alpha}-V_c$ plane, quenched regions can also arise at low circular velocity and there are a large number of star-forming regions at high velocity.  Many outlier spaxels in the $W_{\rm H\alpha}-\beta$ relation represent star-forming regions with declining curves, whereas other are quenched regions with rising curves.

In this section we test whether these outliers are related to the global morphologies of the galaxies in our sample.  Indeed, morphological features like bulges, bars and spiral arms are important galactic dynamical structures (see Section~\ref{S:disc}).  To this end, we examine how the observed spaxels are distributed with respect to several indicators of galactic structure including Hubble type, presence of bar and bulge nature (e.g. classical or pseudo bulge, \citealt{Kormendy2004}; see also Figs. \ref{fig:Vc_stat} and \ref{fig:beta_stat} of this work). 

Following, \citetalias{Kalinova2021} we adopt the Hubble type classifications for the galaxies in our sample from \cite{Walcher2014}.  As discussed in \citetalias{Kalinova2021} (see their Fig. 7), late-type spiral galaxies (Sb$-$Sdm) mostly make up the star-forming class while early-type spirals (Sab$-$Sbc) predominantly make up the quiescent-nuclear-ring and centrally quiescent galaxies. The mixed quenching stage includes a large variety of morphologies (covering early-type spirals, lenticulars and ellipticals).  The nearly and fully retired galaxies consist only of early-type galaxies (both lenticulars and ellipticals). 

According to the bar-type statistics of \citetalias{Kalinova2021}, the quiescent-nuclear-ring group contains the largest number of barred galaxies, followed by centrally quiescent, mixed and star-forming quenching stages.  The nearly- and fully- retired stages, on the other hand, are mostly constituted by unbarred galaxies (see their Fig.8).  It is notable that the quiescent-nuclear-ring, centrally quiescent and mixed quenches stages consist of early-type spiral galaxies (both lenticulars and discy ellipticals).  These systems are thus dominated by secular evolution/dynamical features (e.g. spiral arms and bars; \citetalias{Kalinova2021}), and we expect these galaxies to be the largest contributor to the scatter in the two trends revealed in this work. 

Additionally, we use the central stellar mass surface density of the galaxies within 1 kpc, $\Sigma_{\mathrm{1kpc}}$. This parameter has been used as a proxy for both the mass and compactness of the central region (bulge/spheroid) of the galaxies, scaled on  the same physical size (\citealt{Cheung2012,Luo2020} and references therein).

For each galaxy we define $\Sigma_{\mathrm{1~kpc}}\equiv\frac{M_{*,\mathrm{<1~kpc}}}{\pi\,R^2_{\mathrm{1~kpc}}}$, where M$_{*,\mathrm{<1~kpc}}$ is the stellar mass of the galaxy integrated within the 1 kpc ellipsoid (obtained by de-projecting the radial map on the plane of the galaxy) and R$_{\mathrm{1kpc}}\equiv1$\,kpc (e.g. \citealt{Cheung2012}). For each galaxy, the stellar mass in a given spaxel is determined from the stellar mass surface density map ($\Sigma_{\mathrm{*}}$, included in the PIPE3D dataset; \citealt{Sanchez2016a,Sanchez2016b}) multiplied by the  area of the spaxel. Using $\Sigma_{\mathrm{1kpc}}$, we adopt the method of \cite{Luo2020} to classify the central stellar concentration of the galaxies as either ``classical bulges'' (CB) or ``pseudo bulges'' (PB).  (Elliptical galaxies are omitted from the bulge classification analysis and are designated with the label "E" in the bulge statistics of Figs. \ref{fig:Vc_stat} and \ref{fig:beta_stat}.)  Following \cite{Luo2020}, we use the residual $\Delta\Sigma_{\mathrm{1kpc}}$ from their measured relation between $\Sigma_{\mathrm{1kpc}}$ and global stellar mass (M$_*$, calculated integrating across the entire stellar mass map; equation 2 of \citealt{Luo2020}) to classify galaxies as classical bulge hosts (or elliptical galaxies) when $\log$($\Delta\Sigma_{\mathrm{1kpc}}$) > 0 and as pseudo bulge hosts when $\log$($\Delta\Sigma_{\mathrm{1kpc}}$) < 0 (see \cite{Luo2020}; their Fig. 7). 

As a basic classification of the outliers around the general trend in $W_{\rm{H_{\alpha}}}$ vs. $V_{\rm c}$ (Fig. \ref{fig:Vc_stat}), we define a critical velocity $\log$($V_c$)$\sim$2.4 km s$^{-1}$ ($V_c\sim$ 250 km s$^{-1}$) where the trend changes from relatively flat (in the low $V_c$ region) to steep (in the high $V_c$ region).  Across this critical value, galaxies also separate broadly into star-forming and quenched groups.  Most of the spaxels in low $V_c$ region ($<$ 250 km s$^{-1}$) belong to disc galaxies (S0$-$Sdm) without a dominant bar type (barred, unbarred or unsure) whereas most of the spaxels in the high $V_c$ ($>$ 250 km s$^{-1}$) zone, on the other hand, come from early-type and unbarred galaxies (E1$-$Sab).

The critical velocity also tends to mark a separation in the behaviour of $\Sigma_{\mathrm{1kpc}}$.  For spaxels in the high $V_c$ region (e.g. mainly above 10$^{10}$ M$_\odot$ kpc$^{-2}$), $\Sigma_{\mathrm{1kpc}}$ is also high independent of whether the region is star-forming or quenched. This indicates that regions with high $V_c$ amplitudes are also those with high central stellar mass density.  In the low-$V_c$ regime (e.g. mainly below 10$^{10}$ M$_\odot$ kpc$^{-2}$), in contrast, only quenched regions have high $\Sigma_{\mathrm{1kpc}}$ whereas the star-forming spaxels exhibit a broad bi-modal distribution.  The broadness of the $\Sigma_{\mathrm{1kpc}}$ distribution for this latter subset of galaxies reveals a large variety of central stellar concentrations, and the bi-modality tends to correlate with bulge classification, in the sense that star-forming regions can be arise in galaxies that host both classical bulges (with high $\Sigma_{\mathrm{1kpc}}$) and pseudo bulges (with low $\Sigma_{\mathrm{1kpc}}$). Quenched regions are almost exclusively hosts of classical bulges.

In the $W_{\rm{H_{\alpha}}}-\beta$ diagram, we see that the majority of the spaxels again separate into two main areas: a star-forming$-$ rising CVC zone ($W_{\rm{H_{\alpha}}}$ > 6 {\AA} and $\beta>0$) and a quenched$-$declining CVC region ($W_{\rm{H_{\alpha}}}$ < 3 {\AA} and $\beta < 0$).  In other words, the CVCs of the star-forming galaxies are generally slowly rising ($\beta > 0$), while the CVCs of the quenched systems have a faster inner rise, followed by either a flattened part or a slowly declining part (with $\beta < 0$; outside a central peak caused by a central mass concentration).  The star-forming $-$rising CVC zone is largely populated by spaxels of late-type spirals (Sb$-$Sdm), both barred and unbarred, while the quenched$-$declining CVC region is populated by the spaxels of early-type galaxies (E1$-$Sb), mostly without bars.  This behavior generally agrees with the idea that galactic regions with high stellar mass concentrations (bulges or spheroids) are quenched (e.g. \citealt{Martig2009, Gensior2020}).

There are significant outliers from these general trends, however, with some star-forming galaxies exhibiting declining CVCs ($W_{\rm{H_{\alpha}}}$ > 6 {\AA} and $\beta < 0$) and some quenched galaxies exhibit rising CVCs ($W_{\rm{H_{\alpha}}}$ < 3 {\AA} and $\beta > 0$).  (We note, though, that these outliers represent only a minority of the sample.) Inspection of the individual CVCs in these cases (and across all quenching stages; see Figs. \ref{fig:Vc_EL} and \ref{fig:Vc_EL_flat}), indicates that this quenched subset (of nearly and fully retired class galaxies) are lacking peaks in their CVCs while the subset of star-formers have prominent peaks in their CVCs (e.g., prominent bulges).  The latter are found to arise predominantly in classical bulge hosts.  (In this same zone, quenched regions are also dominated by ellipticals and classical bulges.)  Note, though, that for both star-forming and quenched regions, rising CVCs  are found in galaxies with both classical and pseudo bulges.

Another characteristic of these two outlier regions is that the median of the spaxel distributions is shifted towards early-type disc galaxies (S0$-$Sbc), both with and without bars (Fig. \ref{fig:beta_stat}).  The $\Sigma_{\mathrm{1kpc}}$ spaxel distribution also tends to be tighter compared to the broad distribution characteristic of the star-forming$-$rising CVC zone. 

\section{Concluding remarks}
\label{S:disc}
In this paper, we explore the link between the inner gravitational potential of 215 (E$-$Sdm) non-active CALIFA galaxies through their circular velocity curves (CVC) (calculated from stellar dynamics within 1.5 R$_e$), and the star-formation quenching parameter given by the value of the equivalent width of the H$_{\alpha}$ ($W_{\rm{H_{\alpha}}}$).  
Our main findings can be summarised as follows:

\indent (i) galaxies with certain CVC shape and amplitude tend to be associated with a specific quenching-stage group;\\
\indent (ii) there is a moderate correlation between the amplitude of the velocity of the spaxels ($V_c$) and their quenching-proxy values ($W_{\rm{H_{\alpha}}}$), where the relationship steepens at higher amplitudes above $\log V_c \sim$  2.4 km s$^{-1}$ ($V_c \sim$  250 km s$^{-1}$);\\
\indent (iii) the relation between the shape of the CVC ($\beta$) and  $W_{\rm{H_{\alpha}}}$ is moderate and bi-modal, showing that the quenching regions of the galaxies (where $W_{\rm{H_{\alpha}}}$ is below 3 {\AA}) are overall characterised by high $V_c$ and negative $\beta$ (declining CVC),
and the star-forming regions (where $W_{\rm{H_{\alpha}}}$ is above 6 {\AA}) are overall characterised by low $V_c$ and positive $\beta$ (rising CVC);\\ 
\indent (iv) the outlier spaxels of the $W_{\rm{H_{\alpha}}}-\beta$ relation is largely coming from early-type disc galaxies (S0$-$Sbc) and follows the opposite trend of the main spaxel distribution, described above;\\
\indent (v) the spaxels of the nearly and fully retired class galaxies almost fully occupy the quenched$-$declining CVC region of the $W_{\rm{H_{\alpha}}}-\beta$ relation (where $W_{\rm{H_{\alpha}}}$ < 3 {\AA} and $\beta$ < 0).

These findings indicate an important link of the gravitational potential to the present day star-formation quenching stages of the galaxies. 
The quenching spaxels lay in regions of the galaxies with a large gravitational potential due to a mass concentration (e.g. bulge/spheroid, bar and rings).
Star formation quenching regulated by gravitational potential is generally referred to ``dynamical suppression'' (or ``morphological quenching'' \citealt{Martig2009,Genzel2014, Gensior2020, Gensior2021}). 
Our results support such a scenario, as we find that a proxy for quenching (e.g. $W_{\rm{H_{\alpha}}}$) appears correlated with quantities that trace features of the galactic potential, namely circular velocity curve amplitude ($V_c$) and shape ($\beta$).

A number of other studies have argued that reductions in star formation efficiency (SFE) are responsible for quenching observed in galaxies globally and/or locally (see, e.g. \citealt{Colombo2018,Colombo2020,Ellison2020}). In particular, a link between reduced SFE, morphology, increased circular velocity and rotation curve shear (expressed through the Oort's constant $A=0.5V_{\rm c}/R (1-\beta)$) has been observed by \cite{Colombo2018} within a sample of galaxies with largely similar molecular gas mass surface densities.   That study supports the idea that high shear contributes to the stabilisation of gas discs, preventing the fragmentation in the molecular gas (e.g. \citealt{Toomre1964,Romeo2017}) and reducing the rate of star formation (e.g. \citealt{Martig2009,Meidt2013,Davis2014,Meidt2018}).

Besides shear due to differential rotation (inferred from the CVCs), other kinds of shear may be present due to local gas flows (e.g. \citealt{Meidt2018}), associated either with spirals arms (\citealt{Dobbs_Baba2014,Meidt2013}) or bars (\citealt{Athanassoula1992,Sormani2015}). According to a recent analytical model, local and global (orbital) motions in the galactic potential can lead to a dynamical suppression of star formation on the scales of star-forming giant molecular clouds (\citealt{Meidt2018, Meidt2020,Liu2021}). In this "bottleneck" model, galactic orbital motions compete with gas self-gravity on cloud scales.  These motions contribute to the velocity dispersion within clouds, thus increasing their stable mass (see, e.g. \citealt{Hughes2013} for an observational example of this effect) and introducing environmental variations in gas dynamical state that can influence its ability to collapse and form stars (\citealt{Meidt2013, Leroy2017}).

Whether the reductions in the SFE predicted in some environments by this type of model lead to reductions in the observed rate of star formation depends on the amount of fuel available for star formation. Based on the results of this work, the galaxies most likely to show signs of dynamical suppression are the nearly and fully retired class galaxies in our sample, where the spheroidal component is prominent (e.g. quenched-declining CVC area of the $W_{\rm{H_{\alpha}}}-\beta$ relation, Figs. \ref{fig:Beta_Wha} and \ref{fig:beta_stat}) and gas fractions are low (see also \citealt{Martig2013,Gensior2021}). 
For other systems, dynamical suppression may be less efficient due to the amount and organization of the gas reservoir. This could explain why several of the galaxies in our sample with peaks in their CVCs (indicating a large central mass concentration or bulge) are fully star-forming and why some of the galaxies that are quenched (or are approaching quenching) possess rising curves in our sample. In the first case (i.e. spaxels with $W_{\rm{H_{\alpha}}}$ > 6 {\AA} and $\beta$ < 0), it is noteworthy that galaxies with star-forming classical bulges are less massive and less concentrated than quenched systems with classical bulges (\citealt{Yu2022}), possibly indicating that they have a larger gas reservoir (according to the observed increase in gas fraction with decreasing stellar mass; \citealt{Saintonge2017}). 
Central star formation enhancements in classical bulges (e.g. \citealt{Luo2020}) could also be induced by bars (\citealt{Athanassoula1992, Wang2012, Wang2020}) and spirals (\citealt{Kim2014, Yu2022}). In the second case (i.e. spaxels with $W_{\rm{H_{\alpha}}}$ < 3 {\AA} and $\beta$ > 0), it may be relevant that anaemic spirals and S0s are disk galaxies that can have rising
CVC and no star-formation (\citealt{Kormendy2012}). Therefore, environmental quenching mechanisms may preferentially remove the gas from galaxies with rising CVCs, stopping their star-formation. Spirals could
also fade away as the gas fraction decreases (\citealt{Yu2021}) and becomes S0s (\citealt{Kormendy2012}).

\emph{In summary, our results show that the largest number of spaxels come from  the two extreme quenching stages: star-forming and fully retired classes (corresponding to late-type spirals and elliptical galaxies, respectively) for which  $V_c$  and $\beta$ spaxel distribution coincide with the dynamical suppression hypothesis: the non-quenching  or quenching of the galaxies depends on the absence or presence of bulge/spheroid, respectively (which corresponds to low and high velocities in $W_{\rm{H_{\alpha}}}-V_c$ relation or rising and declining CVC in $W_{\rm{H_{\alpha}}}-\beta$ relation,  respectively). On the other hand, the scatter analysis indicates that the presence of classical bulge is not the only necessary condition. It seems that galaxies need to further have: higher central density (approximately above 10$^{10}$ M$_\odot$ kpc$^{-2}$), no bar, and early-type  morphologies (meaning no tight and prominent spiral arms). }
 
Future work will explore the cold gas content of the studied systems (Kalinova et al., in prep.) and will give a complementary perspective on the dynamics and mechanisms responsible for the formation of the various quenching stages of the nearby galaxies.

\section*{Acknowledgements}
\label{S:acknowledgements}
We thank the anonymous referee, whose suggestions helped us to improve the quality and presentation of this paper.
DC acknowledges support by the \emph{Deut\-sche For\-schungs\-ge\-mein\-schaft, DFG\/} project number SFB956A.
SFS acknowledge Conacyt projects CB-285080 and FC-2016-01 -1916, and PAPIIT-DGAPA IN100519.
RGB and RGD acknowledges financial support from the State Agency for Research of the Spanish MCIU through 
the ``Center of Excellence Severo Ochoa'' award to the Instituto de Astrof\'isica de Andaluc\'ia (SEV-2017-0709), and the projects AYA2016-77846-P and PID2019-109067-GB100.
ER acknowledges the support of the Natural Sciences and Engineering Research Council of Canada (NSERC), funding reference number RGPIN-2017-03987. K.K.  acknowledges  deep  gratitude  to  MPIfR  in  Bonn  for  the Guest Researchership 2013-2021.
 
In this study, we made use of the data of the first legacy survey, 
the Calar Alto Legacy Integral Field Area (CALIFA) survey, 
based on observations made at the Centro Astron\'omico
Hispano Alem\'an (CAHA) at Calar Alto, operated jointly by the Max Planck-Institut
f\"ur Astronomie and the Instituto de Astrof\'isica de Andaluc\'ia
(CSIC).
This research made use of the open-source python packages as \texttt{Astropy} (\citealt{Price2018astropy}), \texttt{SciPy} (\citealt{SciPy2020}), \texttt{NumPy} (\citealt{Harris2020NumPy}), and \texttt{Matplotlib} (\citealt{Hunter2007matplotlib}).


\footnotesize{
\bibliographystyle{aa}
\bibliography{BIB_v9.bib}

\begin{thebibliography}{79}
\expandafter\ifx\csname natexlab\endcsname\relax\def\natexlab#1{#1}\fi

\bibitem[{{Abadi} {et~al.}(1999){Abadi}, {Moore}, \& {Bower}}]{Abadi1999}
{Abadi}, M.~G., {Moore}, B., \& {Bower}, R.~G. 1999, \mnras, 308, 947

\bibitem[{{Alam} {et~al.}(2015){Alam}, {Albareti}, {Allende Prieto}, {Anders},
  {Anderson}, {Anderton}, {Andrews}, {Armengaud}, {Aubourg}, {Bailey}, \&
  et~al.}]{Alam2015}
{Alam}, S., {Albareti}, F.~D., {Allende Prieto}, C., {et~al.} 2015, \apjs, 219,
  12

\bibitem[{{Athanassoula}(1992)}]{Athanassoula1992}
{Athanassoula}, E. 1992, \mnras, 259, 345

\bibitem[{{Balogh} {et~al.}(2000){Balogh}, {Navarro}, \& {Morris}}]{Balogh2000}
{Balogh}, M.~L., {Navarro}, J.~F., \& {Morris}, S.~L. 2000, \apj, 540, 113

\bibitem[{{Belfiore} {et~al.}(2016){Belfiore}, {Maiolino}, {Maraston},
  {Emsellem}, {Bershady}, {Masters}, {Yan}, {Bizyaev}, {Boquien}, {Brownstein},
  {Bundy}, {Drory}, {Heckman}, {Law}, {Roman-Lopes}, {Pan}, {Stanghellini},
  {Thomas}, {Weijmans}, \& {Westfall}}]{Belfiore2016}
{Belfiore}, F., {Maiolino}, R., {Maraston}, C., {et~al.} 2016, \mnras, 461,
  3111

\bibitem[{{Birnboim} \& {Dekel}(2003)}]{Birnboim2003}
{Birnboim}, Y. \& {Dekel}, A. 2003, \mnras, 345, 349

\bibitem[{{Bluck} {et~al.}(2020{\natexlab{a}}){Bluck}, {Maiolino},
  {Piotrowska}, {Trussler}, {Ellison}, {S{\'a}nchez}, {Thorp}, {Teimoorinia},
  {Moreno}, \& {Conselice}}]{Bluck2020b}
{Bluck}, A. F.~L., {Maiolino}, R., {Piotrowska}, J.~M., {et~al.}
  2020{\natexlab{a}}, \mnras, 499, 230

\bibitem[{{Bluck} {et~al.}(2020{\natexlab{b}}){Bluck}, {Maiolino},
  {S{\'a}nchez}, {Ellison}, {Thorp}, {Piotrowska}, {Teimoorinia}, \&
  {Bundy}}]{Bluck2020a}
{Bluck}, A. F.~L., {Maiolino}, R., {S{\'a}nchez}, S.~F., {et~al.}
  2020{\natexlab{b}}, \mnras, 492, 96

\bibitem[{{Cappellari}(2008)}]{Cappellari2008}
{Cappellari}, M. 2008, MNRAS, 390, 71

\bibitem[{{Cheung} {et~al.}(2012){Cheung}, {Faber}, {Koo}, {Dutton}, {Simard},
  {McGrath}, {Huang}, {Bell}, {Dekel}, {Fang}, {Salim}, {Barro}, {Bundy},
  {Coil}, {Cooper}, {Conselice}, {Davis}, {Dom{\'\i}nguez}, {Kassin},
  {Kocevski}, {Koekemoer}, {Lin}, {Lotz}, {Newman}, {Phillips}, {Rosario},
  {Weiner}, \& {Willmer}}]{Cheung2012}
{Cheung}, E., {Faber}, S.~M., {Koo}, D.~C., {et~al.} 2012, \apj, 760, 131

\bibitem[{{Colling} {et~al.}(2018){Colling}, {Hennebelle}, {Geen}, {Iffrig}, \&
  {Bournaud}}]{Colling2018}
{Colling}, C., {Hennebelle}, P., {Geen}, S., {Iffrig}, O., \& {Bournaud}, F.
  2018, \aap, 620, A21

\bibitem[{{Colombo} {et~al.}(2018){Colombo}, {Kalinova}, {Utomo}, {Rosolowsky},
  {Bolatto}, {Levy}, {Wong}, {Sanchez}, {Leroy}, {Ostriker}, {Blitz}, {Vogel},
  {Mast}, {Garc{\'{\i}}a-Benito}, {Husemann}, {Dannerbauer}, {Ellmeier}, \&
  {Cao}}]{Colombo2018}
{Colombo}, D., {Kalinova}, V., {Utomo}, D., {et~al.} 2018, \mnras, 475, 1791

\bibitem[{{Colombo} {et~al.}(2020){Colombo}, {Sanchez}, {Bolatto}, {Kalinova},
  {Wei{\ss}}, {Wong}, {Rosolowsky}, {Vogel}, {Barrera-Ballesteros},
  {Dannerbauer}, {Cao}, {Levy}, {Utomo}, \& {Blitz}}]{Colombo2020}
{Colombo}, D., {Sanchez}, S.~F., {Bolatto}, A.~D., {et~al.} 2020, \aap, 644,
  A97

\bibitem[{{Corcho-Caballero} {et~al.}(2021){Corcho-Caballero}, {Casado},
  {Ascasibar}, \& {Garc{\'\i}a-Benito}}]{Corcho-Caballero2021}
{Corcho-Caballero}, P., {Casado}, J., {Ascasibar}, Y., \& {Garc{\'\i}a-Benito},
  R. 2021, \mnras, 507, 5477

\bibitem[{{Croton} {et~al.}(2006){Croton}, {Springel}, {White}, {De Lucia},
  {Frenk}, {Gao}, {Jenkins}, {Kauffmann}, {Navarro}, \& {Yoshida}}]{Croton2006}
{Croton}, D.~J., {Springel}, V., {White}, S.~D.~M., {et~al.} 2006, \mnras, 365,
  11

\bibitem[{{Davis} {et~al.}(2014){Davis}, {Young}, {Crocker}, {Bureau}, {Blitz},
  {Alatalo}, {Emsellem}, {Naab}, {Bayet}, {Bois}, {Bournaud}, {Cappellari},
  {Davies}, {de Zeeuw}, {Duc}, {Khochfar}, {Krajnovi{\'c}}, {Kuntschner},
  {McDermid}, {Morganti}, {Oosterloo}, {Sarzi}, {Scott}, {Serra}, \&
  {Weijmans}}]{Davis2014}
{Davis}, T.~A., {Young}, L.~M., {Crocker}, A.~F., {et~al.} 2014, \mnras, 444,
  3427

\bibitem[{{Di Matteo} {et~al.}(2005){Di Matteo}, {Springel}, \&
  {Hernquist}}]{DiMatteo2005}
{Di Matteo}, T., {Springel}, V., \& {Hernquist}, L. 2005, \nat, 433, 604

\bibitem[{{Dobbs} \& {Baba}(2014)}]{Dobbs_Baba2014}
{Dobbs}, C. \& {Baba}, J. 2014, \pasa, 31, e035

\bibitem[{{Ellison} {et~al.}(2020){Ellison}, {Thorp}, {Lin}, {Pan}, {Bluck},
  {Scudder}, {Teimoorinia}, {S{\'a}nchez}, \& {Sargent}}]{Ellison2020}
{Ellison}, S.~L., {Thorp}, M.~D., {Lin}, L., {et~al.} 2020, \mnras, 493, L39

\bibitem[{{Emsellem} {et~al.}(1994){Emsellem}, {Monnet}, \&
  {Bacon}}]{Emsellem1994}
{Emsellem}, E., {Monnet}, G., \& {Bacon}, R. 1994, A\&A, 285, 723

\bibitem[{{Faber} {et~al.}(2007){Faber}, {Willmer}, {Wolf}, {Koo}, {Weiner},
  {Newman}, {Im}, {Coil}, {Conroy}, {Cooper}, {Davis}, {Finkbeiner}, {Gerke},
  {Gebhardt}, {Groth}, {Guhathakurta}, {Harker}, {Kaiser}, {Kassin},
  {Kleinheinrich}, {Konidaris}, {Kron}, {Lin}, {Luppino}, {Madgwick},
  {Meisenheimer}, {Noeske}, {Phillips}, {Sarajedini}, {Schiavon}, {Simard},
  {Szalay}, {Vogt}, \& {Yan}}]{Faber2007}
{Faber}, S.~M., {Willmer}, C.~N.~A., {Wolf}, C., {et~al.} 2007, \apj, 665, 265

\bibitem[{{Falc{\'o}n-Barroso} {et~al.}(2017){Falc{\'o}n-Barroso}, {Lyubenova},
  {van de Ven}, {Mendez-Abreu}, {Aguerri}, {Garc{\'{\i}}a-Lorenzo},
  {Bekerait{\'e}}, {S{\'a}nchez}, {Husemann}, {Garc{\'{\i}}a-Benito}, {Mast},
  {Walcher}, {Zibetti}, {Barrera-Ballesteros}, {Galbany},
  {S{\'a}nchez-Bl{\'a}zquez}, {Singh}, {van den Bosch}, {Wild}, {Zhu},
  {Bland-Hawthorn}, {Cid Fernandes}, {de Lorenzo-C{\'a}ceres}, {Gallazzi},
  {Gonz{\'a}lez Delgado}, {Marino}, {M{\'a}rquez}, {P{\'e}rez}, {P{\'e}rez},
  {Roth}, {Rosales-Ortega}, {Ruiz-Lara}, {Wisotzki}, {Ziegler}, \& {Califa
  Collaboration}}]{Falcon-Barroso2017}
{Falc{\'o}n-Barroso}, J., {Lyubenova}, M., {van de Ven}, G., {et~al.} 2017,
  \aap, 597, A48

\bibitem[{{Farouki} \& {Shapiro}(1981)}]{Farouki1981}
{Farouki}, R. \& {Shapiro}, S.~L. 1981, \apj, 243, 32

\bibitem[{{Garc{\'{\i}}a-Benito} {et~al.}(2015){Garc{\'{\i}}a-Benito},
  {Zibetti}, {S{\'a}nchez}, {Husemann}, {de Amorim}, {Castillo-Morales}, {Cid
  Fernandes}, {Ellis}, {Falc{\'o}n-Barroso}, {Galbany}, {Gil de Paz},
  {Gonz{\'a}lez Delgado}, {Lacerda}, {L{\'o}pez-Fernandez}, {de
  Lorenzo-C{\'a}ceres}, {Lyubenova}, {Marino}, {Mast}, {Mendoza}, {P{\'e}rez},
  {Vale Asari}, {Aguerri}, {Ascasibar}, {Bekerait*error*{\.e}},
  {Bland-Hawthorn}, {Barrera-Ballesteros}, {Bomans}, {Cano-D{\'{\i}}az},
  {Catal{\'a}n-Torrecilla}, {Cortijo}, {Delgado-Inglada}, {Demleitner},
  {Dettmar}, {D{\'{\i}}az}, {Florido}, {Gallazzi}, {Garc{\'{\i}}a-Lorenzo},
  {Gomes}, {Holmes}, {Iglesias-P{\'a}ramo}, {Jahnke}, {Kalinova}, {Kehrig},
  {Kennicutt}, {L{\'o}pez-S{\'a}nchez}, {M{\'a}rquez}, {Masegosa}, {Meidt},
  {Mendez-Abreu}, {Moll{\'a}}, {Monreal-Ibero}, {Morisset}, {del Olmo},
  {Papaderos}, {P{\'e}rez}, {Quirrenbach}, {Rosales-Ortega}, {Roth},
  {Ruiz-Lara}, {S{\'a}nchez-Bl{\'a}zquez}, {S{\'a}nchez-Menguiano}, {Singh},
  {Spekkens}, {Stanishev}, {Torres-Papaqui}, {van de Ven}, {Vilchez},
  {Walcher}, {Wild}, {Wisotzki}, {Ziegler}, {Alves}, {Barrado}, {Quintana}, \&
  {Aceituno}}]{Garcia-Benito2015}
{Garc{\'{\i}}a-Benito}, R., {Zibetti}, S., {S{\'a}nchez}, S.~F., {et~al.} 2015,
  \aap, 576, A135

\bibitem[{{Gensior} \& {Kruijssen}(2021)}]{Gensior2021}
{Gensior}, J. \& {Kruijssen}, J.~M.~D. 2021, \mnras, 500, 2000

\bibitem[{{Gensior} {et~al.}(2020){Gensior}, {Kruijssen}, \&
  {Keller}}]{Gensior2020}
{Gensior}, J., {Kruijssen}, J.~M.~D., \& {Keller}, B.~W. 2020, \mnras, 495, 199

\bibitem[{{Genzel} {et~al.}(2014){Genzel}, {F{\"o}rster Schreiber}, {Lang},
  {Tacchella}, {Tacconi}, {Wuyts}, {Bandara}, {Burkert}, {Buschkamp},
  {Carollo}, {Cresci}, {Davies}, {Eisenhauer}, {Hicks}, {Kurk}, {Lilly},
  {Lutz}, {Mancini}, {Naab}, {Newman}, {Peng}, {Renzini}, {Shapiro Griffin},
  {Sternberg}, {Vergani}, {Wisnioski}, {Wuyts}, \& {Zamorani}}]{Genzel2014}
{Genzel}, R., {F{\"o}rster Schreiber}, N.~M., {Lang}, P., {et~al.} 2014, \apj,
  785, 75

\bibitem[{{Gunn} \& {Gott}(1972)}]{Gunn1972}
{Gunn}, J.~E. \& {Gott}, J.~Richard, I. 1972, \apj, 176, 1

\bibitem[{Harris {et~al.}(2020)Harris, Millman, van~der Walt, Gommers,
  Virtanen, Cournapeau, Wieser, Taylor, Berg, Smith, Kern, Picus, Hoyer, van
  Kerkwijk, Brett, Haldane, del R{\'{i}}o, Wiebe, Peterson,
  G{\'{e}}rard-Marchant, Sheppard, Reddy, Weckesser, Abbasi, Gohlke, \&
  Oliphant}]{Harris2020NumPy}
Harris, C.~R., Millman, K.~J., van~der Walt, S.~J., {et~al.} 2020, Nature, 585,
  357

\bibitem[{{Hughes} {et~al.}(2013){Hughes}, {Meidt}, {Colombo}, {Schinnerer},
  {Pety}, {Leroy}, {Dobbs}, {Garc{\'\i}a-Burillo}, {Thompson}, {Dumas},
  {Schuster}, \& {Kramer}}]{Hughes2013}
{Hughes}, A., {Meidt}, S.~E., {Colombo}, D., {et~al.} 2013, \apj, 779, 46

\bibitem[{Hunter(2007)}]{Hunter2007matplotlib}
Hunter, J.~D. 2007, Computing in science \& engineering, 9, 90

\bibitem[{{Husemann} \& {Harrison}(2018)}]{Husemann2018}
{Husemann}, B. \& {Harrison}, C.~M. 2018, Nature Astronomy, 2, 196

\bibitem[{{Husemann} {et~al.}(2013){Husemann}, {Jahnke}, {S{\'a}nchez},
  {Barrado}, {Bekerait*error*{\.e}}, {Bomans}, {Castillo-Morales},
  {Catal{\'a}n-Torrecilla}, {Cid Fernandes}, {Falc{\'o}n-Barroso},
  {Garc{\'{\i}}a-Benito}, {Gonz{\'a}lez Delgado}, {Iglesias-P{\'a}ramo},
  {Johnson}, {Kupko}, {L{\'o}pez-Fernandez}, {Lyubenova}, {Marino}, {Mast},
  {Miskolczi}, {Monreal-Ibero}, {Gil de Paz}, {P{\'e}rez}, {P{\'e}rez},
  {Rosales-Ortega}, {Ruiz-Lara}, {Schilling}, {van de Ven}, {Walcher}, {Alves},
  {de Amorim}, {Backsmann}, {Barrera-Ballesteros}, {Bland-Hawthorn}, {Cortijo},
  {Dettmar}, {Demleitner}, {D{\'{\i}}az}, {Enke}, {Florido}, {Flores},
  {Galbany}, {Gallazzi}, {Garc{\'{\i}}a-Lorenzo}, {Gomes}, {Gruel}, {Haines},
  {Holmes}, {Jungwiert}, {Kalinova}, {Kehrig}, {Kennicutt}, {Klar}, {Lehnert},
  {L{\'o}pez-S{\'a}nchez}, {de Lorenzo-C{\'a}ceres}, {M{\'a}rmol-Queralt{\'o}},
  {M{\'a}rquez}, {Mendez-Abreu}, {Moll{\'a}}, {del Olmo}, {Meidt}, {Papaderos},
  {Puschnig}, {Quirrenbach}, {Roth}, {S{\'a}nchez-Bl{\'a}zquez}, {Spekkens},
  {Singh}, {Stanishev}, {Trager}, {Vilchez}, {Wild}, {Wisotzki}, {Zibetti}, \&
  {Ziegler}}]{Husemann2013}
{Husemann}, B., {Jahnke}, K., {S{\'a}nchez}, S.~F., {et~al.} 2013, \aap, 549,
  A87

\bibitem[{{Kalinova} {et~al.}(2017){Kalinova}, {Colombo}, {Rosolowsky},
  {Kannan}, {Galbany}, {Garc{\'{\i}}a-Benito}, {Gonz{\'a}lez Delgado},
  {S{\'a}nchez}, {Ruiz-Lara}, {M{\'e}ndez-Abreu}, {Catal{\'a}n-Torrecilla},
  {S{\'a}nchez-Menguiano}, {de Lorenzo-C{\'a}ceres}, {Costantin}, {Florido},
  {Kodaira}, {Marino}, {L{\"a}sker}, \& {Bland-Hawthorn}}]{Kalinova2017b}
{Kalinova}, V., {Colombo}, D., {Rosolowsky}, E., {et~al.} 2017, \mnras, 469,
  2539

\bibitem[{{Kalinova} {et~al.}(2021){Kalinova}, {Colombo}, {S{\'a}nchez},
  {Kodaira}, {Garc{\'\i}a-Benito}, {Gonz{\'a}lez Delgado}, {Rosolowsky}, \&
  {Lacerda}}]{Kalinova2021}
{Kalinova}, V., {Colombo}, D., {S{\'a}nchez}, S.~F., {et~al.} 2021, \aap, 648,
  A64

\bibitem[{{Khoperskov} {et~al.}(2018){Khoperskov}, {Haywood}, {Di Matteo},
  {Lehnert}, \& {Combes}}]{Khoperskov2018}
{Khoperskov}, S., {Haywood}, M., {Di Matteo}, P., {Lehnert}, M.~D., \&
  {Combes}, F. 2018, \aap, 609, A60

\bibitem[{{Kim} \& {Kim}(2014)}]{Kim2014}
{Kim}, Y. \& {Kim}, W.-T. 2014, \mnras, 440, 208

\bibitem[{{Kormendy} \& {Bender}(2012)}]{Kormendy2012}
{Kormendy}, J. \& {Bender}, R. 2012, \apjs, 198, 2

\bibitem[{{Kormendy} \& {Kennicutt}(2004)}]{Kormendy2004}
{Kormendy}, J. \& {Kennicutt}, Jr., R.~C. 2004, \araa, 42, 603

\bibitem[{Kowalski(1972)}]{Kowalski1972}
Kowalski, C.~J. 1972, Journal of the Royal Statistical Society. Series C
  (Applied Statistics), 21, 1

\bibitem[{{Krumholz} \& {Kruijssen}(2015)}]{Krumholz2015}
{Krumholz}, M.~R. \& {Kruijssen}, J.~M.~D. 2015, \mnras, 453, 739

\bibitem[{{Lacerda} {et~al.}(2018){Lacerda}, {Cid Fernandes}, {Couto},
  {Stasi{\'n}ska}, {Garc{\'{\i}}a-Benito}, {Vale Asari}, {P{\'e}rez},
  {Gonz{\'a}lez Delgado}, {S{\'a}nchez}, \& {de Amorim}}]{Lacerda2018}
{Lacerda}, E.~A.~D., {Cid Fernandes}, R., {Couto}, G.~S., {et~al.} 2018,
  \mnras, 474, 3727

\bibitem[{{Lacerda} {et~al.}(2020){Lacerda}, {S{\'a}nchez}, {Cid Fernandes},
  {L{\'o}pez-Cob{\'a}}, {Espinosa-Ponce}, \& {Galbany}}]{Lacerda2020}
{Lacerda}, E. A.~D., {S{\'a}nchez}, S.~F., {Cid Fernandes}, R., {et~al.} 2020,
  \mnras, 492, 3073

\bibitem[{{Larson} {et~al.}(1980){Larson}, {Tinsley}, \&
  {Caldwell}}]{Larson1980}
{Larson}, R.~B., {Tinsley}, B.~M., \& {Caldwell}, C.~N. 1980, \apj, 237, 692

\bibitem[{{Leroy} {et~al.}(2017){Leroy}, {Schinnerer}, {Hughes}, {Kruijssen},
  {Meidt}, {Schruba}, {Sun}, {Bigiel}, {Aniano}, {Blanc}, {Bolatto},
  {Chevance}, {Colombo}, {Gallagher}, {Garcia-Burillo}, {Kramer}, {Querejeta},
  {Pety}, {Thompson}, \& {Usero}}]{Leroy2017}
{Leroy}, A.~K., {Schinnerer}, E., {Hughes}, A., {et~al.} 2017, \apj, 846, 71

\bibitem[{{Leroy} {et~al.}(2008){Leroy}, {Walter}, {Brinks}, {Bigiel}, {de
  Blok}, {Madore}, \& {Thornley}}]{Leroy2008}
{Leroy}, A.~K., {Walter}, F., {Brinks}, E., {et~al.} 2008, \aj, 136, 2782

\bibitem[{{Liu} {et~al.}(2021){Liu}, {Bureau}, {Blitz}, {Davis}, {Onishi},
  {Smith}, {North}, \& {Iguchi}}]{Liu2021}
{Liu}, L., {Bureau}, M., {Blitz}, L., {et~al.} 2021, \mnras, 505, 4048

\bibitem[{{Luo} {et~al.}(2020){Luo}, {Faber}, {Rodr{\'\i}guez-Puebla}, {Woo},
  {Guo}, {Koo}, {Primack}, {Chen}, {Yesuf}, {Lin}, {Barro}, {Fang}, {Pandya},
  {Huertas-Company}, \& {Mao}}]{Luo2020}
{Luo}, Y., {Faber}, S.~M., {Rodr{\'\i}guez-Puebla}, A., {et~al.} 2020, \mnras,
  493, 1686

\bibitem[{{Martig} {et~al.}(2009){Martig}, {Bournaud}, {Teyssier}, \&
  {Dekel}}]{Martig2009}
{Martig}, M., {Bournaud}, F., {Teyssier}, R., \& {Dekel}, A. 2009, \apj, 707,
  250

\bibitem[{{Martig} {et~al.}(2013){Martig}, {Crocker}, {Bournaud}, {Emsellem},
  {Gabor}, {Alatalo}, {Blitz}, {Bois}, {Bureau}, {Cappellari}, {Davies},
  {Davis}, {Dekel}, {de Zeeuw}, {Duc}, {Falc{\'o}n-Barroso}, {Khochfar},
  {Krajnovi{\'c}}, {Kuntschner}, {Morganti}, {McDermid}, {Naab}, {Oosterloo},
  {Sarzi}, {Scott}, {Serra}, {Griffin}, {Teyssier}, {Weijmans}, \&
  {Young}}]{Martig2013}
{Martig}, M., {Crocker}, A.~F., {Bournaud}, F., {et~al.} 2013, \mnras, 432,
  1914

\bibitem[{{Meidt} {et~al.}(2020){Meidt}, {Glover}, {Kruijssen}, {Leroy},
  {Rosolowsky}, {Hughes}, {Schinnerer}, {Schruba}, {Usero}, {Bigiel}, {Blanc},
  {Chevance}, {Pety}, {Querejeta}, \& {Utomo}}]{Meidt2020}
{Meidt}, S.~E., {Glover}, S. C.~O., {Kruijssen}, J.~M.~D., {et~al.} 2020, \apj,
  892, 73

\bibitem[{{Meidt} {et~al.}(2018){Meidt}, {Leroy}, {Rosolowsky}, {Kruijssen},
  {Schinnerer}, {Schruba}, {Pety}, {Blanc}, {Bigiel}, {Chevance}, {Hughes},
  {Querejeta}, \& {Usero}}]{Meidt2018}
{Meidt}, S.~E., {Leroy}, A.~K., {Rosolowsky}, E., {et~al.} 2018, \apj, 854, 100

\bibitem[{{Meidt} {et~al.}(2013){Meidt}, {Schinnerer}, {Garc{\'{\i}}a-Burillo},
  {Hughes}, {Colombo}, {Pety}, {Dobbs}, {Schuster}, {Kramer}, {Leroy}, {Dumas},
  \& {Thompson}}]{Meidt2013}
{Meidt}, S.~E., {Schinnerer}, E., {Garc{\'{\i}}a-Burillo}, S., {et~al.} 2013,
  \apj, 779, 45

\bibitem[{{Monnet} {et~al.}(1992){Monnet}, {Bacon}, \& {Emsellem}}]{Monnet1992}
{Monnet}, G., {Bacon}, R., \& {Emsellem}, E. 1992, A\&A, 253, 366

\bibitem[{{Moore} {et~al.}(1996){Moore}, {Katz}, {Lake}, {Dressler}, \&
  {Oemler}}]{Moore1996}
{Moore}, B., {Katz}, N., {Lake}, G., {Dressler}, A., \& {Oemler}, A. 1996,
  \nat, 379, 613

\bibitem[{{Peng} {et~al.}(2010){Peng}, {Lilly}, {Kova{\v{c}}}, {Bolzonella},
  {Pozzetti}, {Renzini}, {Zamorani}, {Ilbert}, {Knobel}, {Iovino}, {Maier},
  {Cucciati}, {Tasca}, {Carollo}, {Silverman}, {Kampczyk}, {de Ravel},
  {Sanders}, {Scoville}, {Contini}, {Mainieri}, {Scodeggio}, {Kneib}, {Le
  F{\`e}vre}, {Bardelli}, {Bongiorno}, {Caputi}, {Coppa}, {de la Torre},
  {Franzetti}, {Garilli}, {Lamareille}, {Le Borgne}, {Le Brun}, {Mignoli},
  {Perez Montero}, {Pello}, {Ricciardelli}, {Tanaka}, {Tresse}, {Vergani},
  {Welikala}, {Zucca}, {Oesch}, {Abbas}, {Barnes}, {Bordoloi}, {Bottini},
  {Cappi}, {Cassata}, {Cimatti}, {Fumana}, {Hasinger}, {Koekemoer},
  {Leauthaud}, {Maccagni}, {Marinoni}, {McCracken}, {Memeo}, {Meneux}, {Nair},
  {Porciani}, {Presotto}, \& {Scaramella}}]{Peng2010}
{Peng}, Y.-j., {Lilly}, S.~J., {Kova{\v{c}}}, K., {et~al.} 2010, \apj, 721, 193

\bibitem[{Price-Whelan {et~al.}(2018)Price-Whelan, Sip{\H{o}}cz, G{\"u}nther,
  Lim, Crawford, Conseil, Shupe, Craig, Dencheva, Ginsburg,
  {et~al.}}]{Price2018astropy}
Price-Whelan, A.~M., Sip{\H{o}}cz, B., G{\"u}nther, H., {et~al.} 2018, The
  Astronomical Journal, 156, 123

\bibitem[{{Romeo} \& {Fathi}(2015)}]{Romeo2015}
{Romeo}, A.~B. \& {Fathi}, K. 2015, \mnras, 451, 3107

\bibitem[{{Romeo} \& {Fathi}(2016)}]{Romeo2016}
{Romeo}, A.~B. \& {Fathi}, K. 2016, \mnras, 460, 2360

\bibitem[{{Romeo} \& {Mogotsi}(2017)}]{Romeo2017}
{Romeo}, A.~B. \& {Mogotsi}, K.~M. 2017, \mnras, 469, 286

\bibitem[{{Saintonge} {et~al.}(2017){Saintonge}, {Catinella}, {Tacconi},
  {Kauffmann}, {Genzel}, {Cortese}, {Dav{\'e}}, {Fletcher},
  {Graci{\'a}-Carpio}, {Kramer}, {Heckman}, {Janowiecki}, {Lutz}, {Rosario},
  {Schiminovich}, {Schuster}, {Wang}, {Wuyts}, {Borthakur}, {Lamperti}, \&
  {Roberts-Borsani}}]{Saintonge2017}
{Saintonge}, A., {Catinella}, B., {Tacconi}, L.~J., {et~al.} 2017, \apjs, 233,
  22

\bibitem[{{Salpeter}(1955)}]{Salpeter1955}
{Salpeter}, E.~E. 1955, A\&A, 121, 161

\bibitem[{{S{\'a}nchez} {et~al.}(2016{\natexlab{a}}){S{\'a}nchez},
  {Garc{\'{\i}}a-Benito}, {Zibetti}, {Walcher}, {Husemann}, {Mendoza},
  {Galbany}, {Falc{\'o}n-Barroso}, {Mast}, {Aceituno}, {Aguerri}, {Alves},
  {Amorim}, {Ascasibar}, {Barrado-Navascues}, {Barrera-Ballesteros},
  {Bekerait{\`e}}, {Bland-Hawthorn}, {Cano D{\'{\i}}az}, {Cid Fernandes},
  {Cavichia}, {Cortijo}, {Dannerbauer}, {Demleitner}, {D{\'{\i}}az}, {Dettmar},
  {de Lorenzo-C{\'a}ceres}, {del Olmo}, {Galazzi}, {Garc{\'{\i}}a-Lorenzo},
  {Gil de Paz}, {Gonz{\'a}lez Delgado}, {Holmes}, {Igl{\'e}sias-P{\'a}ramo},
  {Kehrig}, {Kelz}, {Kennicutt}, {Kleemann}, {Lacerda}, {L{\'o}pez
  Fern{\'a}ndez}, {L{\'o}pez S{\'a}nchez}, {Lyubenova}, {Marino},
  {M{\'a}rquez}, {Mendez-Abreu}, {Moll{\'a}}, {Monreal-Ibero}, {Ortega
  Minakata}, {Torres-Papaqui}, {P{\'e}rez}, {Rosales-Ortega}, {Roth},
  {S{\'a}nchez-Bl{\'a}zquez}, {Schilling}, {Spekkens}, {Vale Asari}, {van den
  Bosch}, {van de Ven}, {Vilchez}, {Wild}, {Wisotzki}, {Y{\i}ld{\i}r{\i}m}, \&
  {Ziegler}}]{Sanchez2016c}
{S{\'a}nchez}, S.~F., {Garc{\'{\i}}a-Benito}, R., {Zibetti}, S., {et~al.}
  2016{\natexlab{a}}, \aap, 594, A36

\bibitem[{{S{\'a}nchez} {et~al.}(2012){S{\'a}nchez}, {Kennicutt}, {Gil de Paz},
  {van de Ven}, {V{\'{\i}}lchez}, {Wisotzki}, {Walcher}, {Mast}, {Aguerri},
  {Albiol-P{\'e}rez}, {Alonso-Herrero}, {Alves}, {Bakos}, {Bart{\'a}kov{\'a}},
  {Bland-Hawthorn}, {Boselli}, {Bomans}, {Castillo-Morales}, {Cortijo-Ferrero},
  {de Lorenzo-C{\'a}ceres}, {Del Olmo}, {Dettmar}, {D{\'{\i}}az}, {Ellis},
  {Falc{\'o}n-Barroso}, {Flores}, {Gallazzi}, {Garc{\'{\i}}a-Lorenzo},
  {Gonz{\'a}lez Delgado}, {Gruel}, {Haines}, {Hao}, {Husemann},
  {Igl{\'e}sias-P{\'a}ramo}, {Jahnke}, {Johnson}, {Jungwiert}, {Kalinova},
  {Kehrig}, {Kupko}, {L{\'o}pez-S{\'a}nchez}, {Lyubenova}, {Marino},
  {M{\'a}rmol-Queralt{\'o}}, {M{\'a}rquez}, {Masegosa}, {Meidt},
  {Mendez-Abreu}, {Monreal-Ibero}, {Montijo}, {Mour{\~a}o}, {Palacios-Navarro},
  {Papaderos}, {Pasquali}, {Peletier}, {P{\'e}rez}, {P{\'e}rez}, {Quirrenbach},
  {Rela{\~n}o}, {Rosales-Ortega}, {Roth}, {Ruiz-Lara},
  {S{\'a}nchez-Bl{\'a}zquez}, {Sengupta}, {Singh}, {Stanishev}, {Trager},
  {Vazdekis}, {Viironen}, {Wild}, {Zibetti}, \& {Ziegler}}]{Sanchez2012}
{S{\'a}nchez}, S.~F., {Kennicutt}, R.~C., {Gil de Paz}, A., {et~al.} 2012,
  \aap, 538, A8

\bibitem[{{S{\'a}nchez} {et~al.}(2016{\natexlab{b}}){S{\'a}nchez}, {P{\'e}rez},
  {S{\'a}nchez-Bl{\'a}zquez}, {Garc{\'\i}a-Benito}, {Ibarra-Mede},
  {Gonz{\'a}lez}, {Rosales-Ortega}, {S{\'a}nchez-Menguiano}, {Ascasibar},
  {Bitsakis}, {Law}, {Cano-D{\'\i}az}, {L{\'o}pez-Cob{\'a}}, {Marino}, {Gil de
  Paz}, {L{\'o}pez-S{\'a}nchez}, {Barrera-Ballesteros}, {Galbany}, {Mast},
  {Abril-Melgarejo}, \& {Roman-Lopes}}]{Sanchez2016b}
{S{\'a}nchez}, S.~F., {P{\'e}rez}, E., {S{\'a}nchez-Bl{\'a}zquez}, P., {et~al.}
  2016{\natexlab{b}}, \rmxaa, 52, 171

\bibitem[{{S{\'a}nchez} {et~al.}(2016{\natexlab{c}}){S{\'a}nchez}, {P{\'e}rez},
  {S{\'a}nchez-Bl{\'a}zquez}, {Gonz{\'a}lez}, {Ros{\'a}lez-Ortega},
  {Cano-D{\'\i} az}, {L{\'o}pez-Cob{\'a}}, {Marino}, {Gil de Paz}, {Moll{\'a}},
  {L{\'o}pez-S{\'a}nchez}, {Ascasibar}, \&
  {Barrera-Ballesteros}}]{Sanchez2016a}
{S{\'a}nchez}, S.~F., {P{\'e}rez}, E., {S{\'a}nchez-Bl{\'a}zquez}, P., {et~al.}
  2016{\natexlab{c}}, \rmxaa, 52, 21

\bibitem[{{S{\'a}nchez} {et~al.}(2014){S{\'a}nchez}, {Rosales-Ortega},
  {Iglesias-P{\'a}ramo}, {Moll{\'a}}, {Barrera-Ballesteros}, {Marino},
  {P{\'e}rez}, {S{\'a}nchez-Blazquez}, {Gonz{\'a}lez Delgado}, {Cid Fernand
  es}, {de Lorenzo-C{\'a}ceres}, {Mendez-Abreu}, {Galbany}, {Falcon-Barroso},
  {Miralles-Caballero}, {Husemann}, {Garc{\'\i}a-Benito}, {Mast}, {Walcher},
  {Gil de Paz}, {Garc{\'\i}a-Lorenzo}, {Jungwiert}, {V{\'\i}lchez},
  {J{\'\i}lkov{\'a}}, {Lyubenova}, {Cortijo-Ferrero}, {D{\'\i}az}, {Wisotzki},
  {M{\'a}rquez}, {Bland-Hawthorn}, {Ellis}, {van de Ven}, {Jahnke},
  {Papaderos}, {Gomes}, {Mendoza}, \& {L{\'o}pez-S{\'a}nchez}}]{Sanchez2014}
{S{\'a}nchez}, S.~F., {Rosales-Ortega}, F.~F., {Iglesias-P{\'a}ramo}, J.,
  {et~al.} 2014, \aap, 563, A49

\bibitem[{Savitzky \& Golay(1964)}]{Savitzky-Golay1964}
Savitzky, A. \& Golay, M. J.~E. 1964, Analytical Chemistry, 36, 1627

\bibitem[{{Singh} {et~al.}(2013){Singh}, {van de Ven}, {Jahnke}, {Lyubenova},
  {Falc{\'o}n-Barroso}, {Alves}, {Cid Fernandes}, {Galbany},
  {Garc{\'\i}a-Benito}, {Husemann}, {Kennicutt}, {Marino}, {M{\'a}rquez},
  {Masegosa}, {Mast}, {Pasquali}, {S{\'a}nchez}, {Walcher}, {Wild}, {Wisotzki},
  \& {Ziegler}}]{Singh2013}
{Singh}, R., {van de Ven}, G., {Jahnke}, K., {et~al.} 2013, \aap, 558, A43

\bibitem[{{Sormani} {et~al.}(2015){Sormani}, {Binney}, \&
  {Magorrian}}]{Sormani2015}
{Sormani}, M.~C., {Binney}, J., \& {Magorrian}, J. 2015, \mnras, 449, 2421

\bibitem[{{Strateva} {et~al.}(2001){Strateva}, {Ivezi{\'c}}, {Knapp},
  {Narayanan}, {Strauss}, {Gunn}, {Lupton}, {Schlegel}, {Bahcall}, {Brinkmann},
  {Brunner}, {Budav{\'a}ri}, {Csabai}, {Castander}, {Doi}, {Fukugita},
  {Gy{\H{o}}ry}, {Hamabe}, {Hennessy}, {Ichikawa}, {Kunszt}, {Lamb}, {McKay},
  {Okamura}, {Racusin}, {Sekiguchi}, {Schneider}, {Shimasaku}, \&
  {York}}]{Strateva2001}
{Strateva}, I., {Ivezi{\'c}}, {\v{Z}}., {Knapp}, G.~R., {et~al.} 2001, \aj,
  122, 1861

\bibitem[{{Toomre}(1964)}]{Toomre1964}
{Toomre}, A. 1964, \apj, 139, 1217

\bibitem[{Virtanen {et~al.}(2020)Virtanen, Gommers, Oliphant, Haberland, Reddy,
  Cournapeau, Burovski, Peterson, Weckesser, Bright, {van der Walt}, Brett,
  Wilson, Millman, Mayorov, Nelson, Jones, Kern, Larson, Carey, Polat, Feng,
  Moore, {VanderPlas}, Laxalde, Perktold, Cimrman, Henriksen, Quintero, Harris,
  Archibald, Ribeiro, Pedregosa, {van Mulbregt}, \& {SciPy 1.0
  Contributors}}]{SciPy2020}
Virtanen, P., Gommers, R., Oliphant, T.~E., {et~al.} 2020, Nature Methods, 17,
  261

\bibitem[{{Walcher} {et~al.}(2014){Walcher}, {Wisotzki}, {Bekerait{\'e}},
  {Husemann}, {Iglesias-P{\'a}ramo}, {Backsmann}, {Barrera Ballesteros},
  {Catal{\'a}n-Torrecilla}, {Cortijo}, {del Olmo}, {Garcia Lorenzo},
  {Falc{\'o}n-Barroso}, {Jilkova}, {Kalinova}, {Mast}, {Marino},
  {M{\'e}ndez-Abreu}, {Pasquali}, {S{\'a}nchez}, {Trager}, {Zibetti},
  {Aguerri}, {Alves}, {Bland-Hawthorn}, {Boselli}, {Castillo Morales}, {Cid
  Fernandes}, {Flores}, {Galbany}, {Gallazzi}, {Garc{\'{\i}}a-Benito}, {Gil de
  Paz}, {Gonz{\'a}lez-Delgado}, {Jahnke}, {Jungwiert}, {Kehrig}, {Lyubenova},
  {M{\'a}rquez Perez}, {Masegosa}, {Monreal Ibero}, {P{\'e}rez}, {Quirrenbach},
  {Rosales-Ortega}, {Roth}, {Sanchez-Blazquez}, {Spekkens}, {Tundo}, {van de
  Ven}, {Verheijen}, {Vilchez}, \& {Ziegler}}]{Walcher2014}
{Walcher}, C.~J., {Wisotzki}, L., {Bekerait{\'e}}, S., {et~al.} 2014, \aap,
  569, A1

\bibitem[{{Wang} {et~al.}(2020){Wang}, {Athanassoula}, {Yu}, {Wolf}, {Shao},
  {Gao}, \& {Randriamampandry}}]{Wang2020}
{Wang}, J., {Athanassoula}, E., {Yu}, S.-Y., {et~al.} 2020, \apj, 893, 19

\bibitem[{{Wang} {et~al.}(2012){Wang}, {Kauffmann}, {Overzier}, {Tacconi},
  {Kong}, {Saintonge}, {Catinella}, {Schiminovich}, {Moran}, \&
  {Johnson}}]{Wang2012}
{Wang}, J., {Kauffmann}, G., {Overzier}, R., {et~al.} 2012, \mnras, 423, 3486

\bibitem[{{Yu} {et~al.}(2021){Yu}, {Ho}, \& {Wang}}]{Yu2021}
{Yu}, S.-Y., {Ho}, L.~C., \& {Wang}, J. 2021, \apj, 917, 88

\bibitem[{{Yu} {et~al.}(2022){Yu}, {Xu}, {Ho}, {Wang}, \& {Kao}}]{Yu2022}
{Yu}, S.-Y., {Xu}, D., {Ho}, L.~C., {Wang}, J., \& {Kao}, W.-B. 2022, \aap,
  661, A98

\bibitem[{Zwillinger \& Kokoska(1999)}]{Zwillinger1999}
Zwillinger, D. \& Kokoska, S. 1999, CRC Standard Probability and Statistics
  Tables and Formulae (CRC Press)

\end{thebibliography}
}

 \appendix
\section{Circular velocity curves across quenching stage}
\label{A:cvc_all}
\begin{figure*}
\centering
\includegraphics[width=0.85\textwidth]{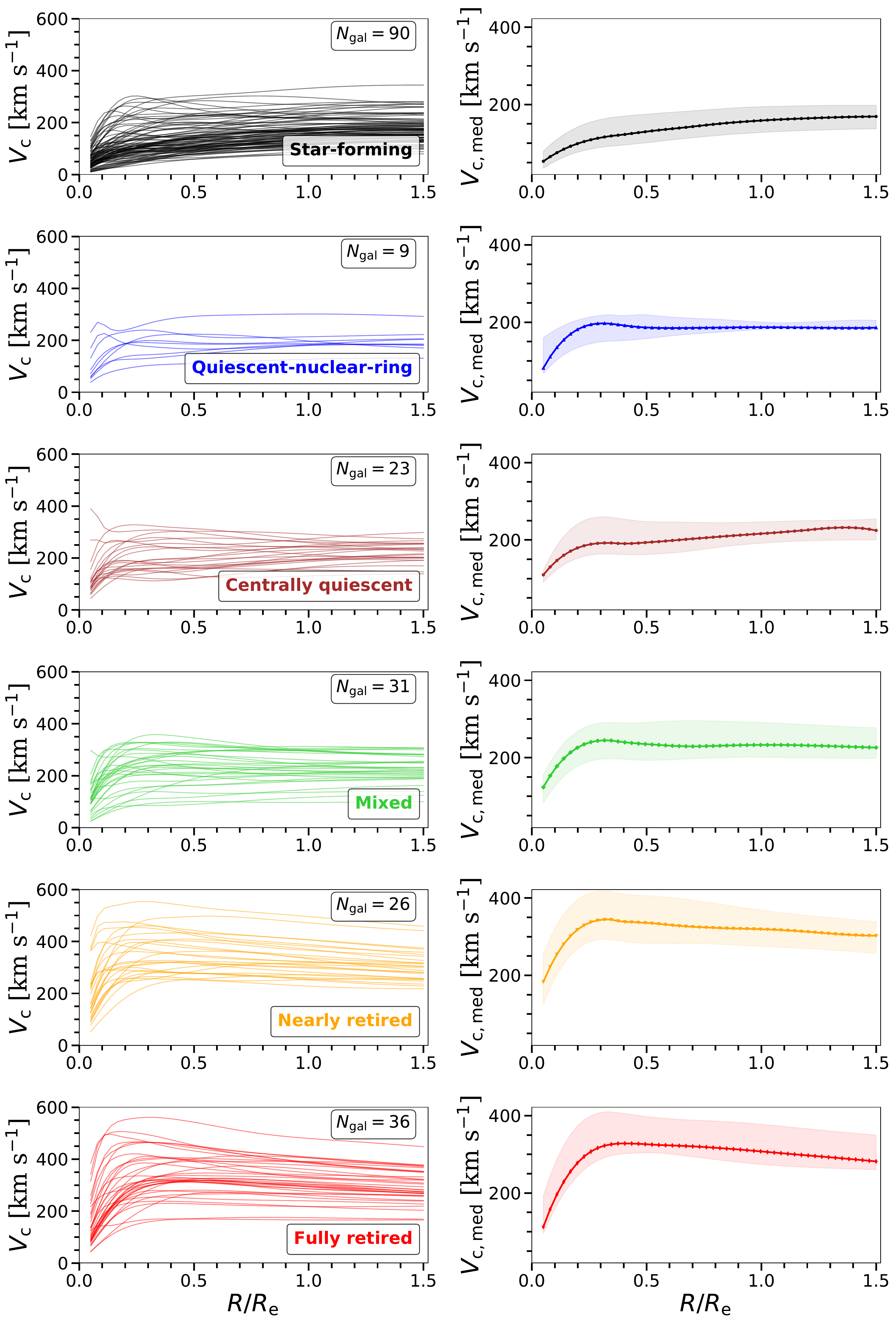}
\caption{Circular velocity curves of our sample (thin colour lines in the left row panels), grouped in the six quenching stages as defined  in \citetalias{Kalinova2021}.
The median circular velocity curve (thick colour lines) for each quenching group is shown in the right row panels. } 
\label{fig:Vc_EL}
\end{figure*}

\begin{figure*}
\centering
\includegraphics[width=0.85\textwidth]{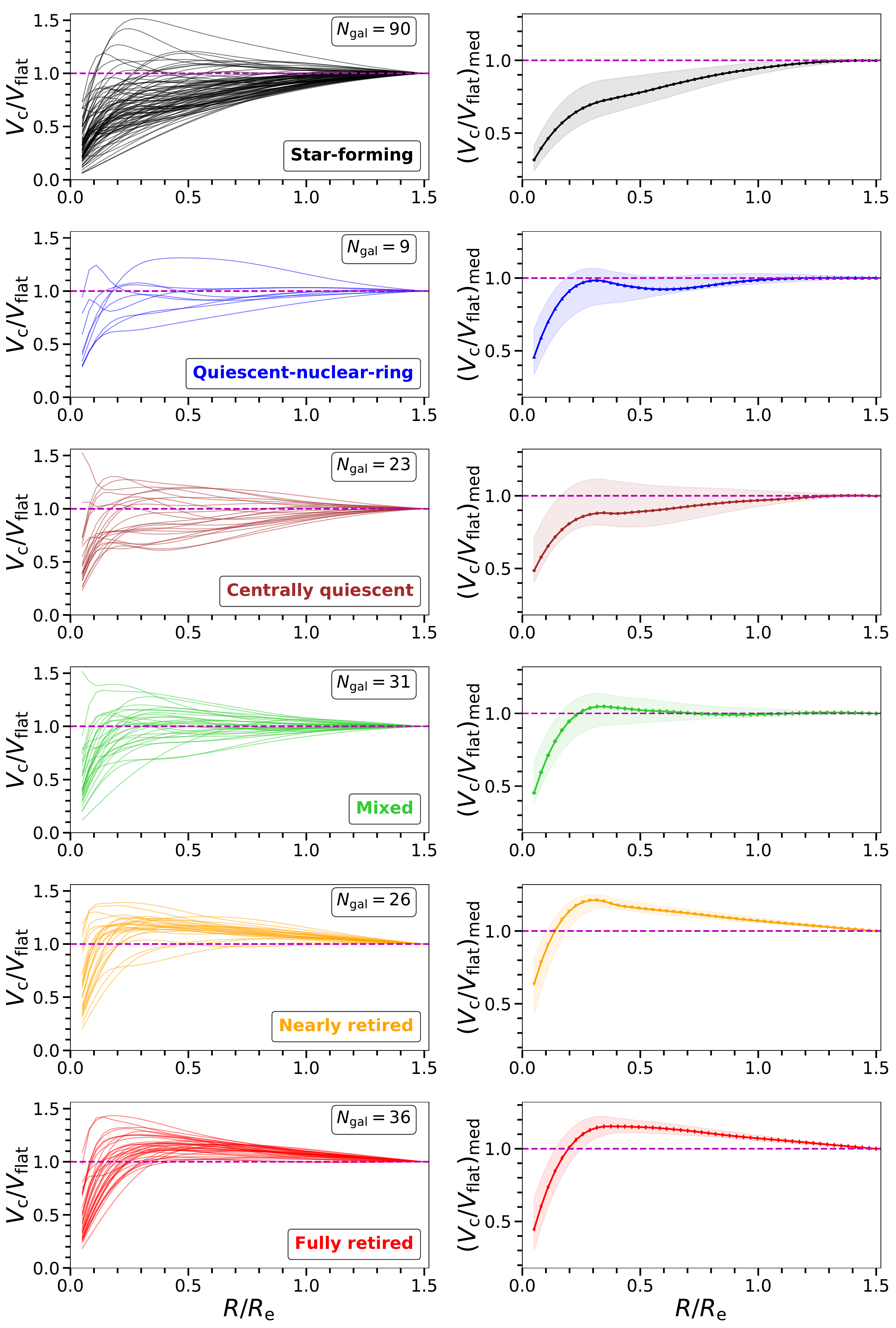}
\caption{Normalised on the asymptotic speed circular velocity curves of our sample (thin colour lines in the left row panels), grouped in the six quenching stages as defined  in \citetalias{Kalinova2021}.
The median smoothed curve of the normalised on the asymptotic speeds circular velocity curves (thick colour lines) for each quenching group is shown in the right row panels. The shadowed regions indicate the dispersion of the curves from the median value. The magenta horizontal line indicates the unity value between the two velocities. } 
\label{fig:Vc_EL_flat}
\end{figure*}

In the left panels of Figs. \ref{fig:Vc_EL}  and \ref{fig:Vc_EL_flat}, we show the individual circular velocity curves (CVCs) of the galaxies and the normalised on their asymptotic velocity CVCs per quenching groups, respectively.
The right panels of the figures present the median and percentile (25th and 75th) profiles of the CVCs. The median CVCs and their percentile profiles are smoothed through  the Savitzky-Golay smoothing filter (\citealt{Savitzky-Golay1964}) by adopting third degree polynomial and 21-point wide sliding window.

\section{Individual $W_{\rm{H_{\alpha}}} - V_c$ and $W_{\rm{H_{\alpha}}} - \beta$ distribution per quenching stage}
\label{A:resolved_each}
\begin{figure*}
\centering
\includegraphics[width=1\textwidth]{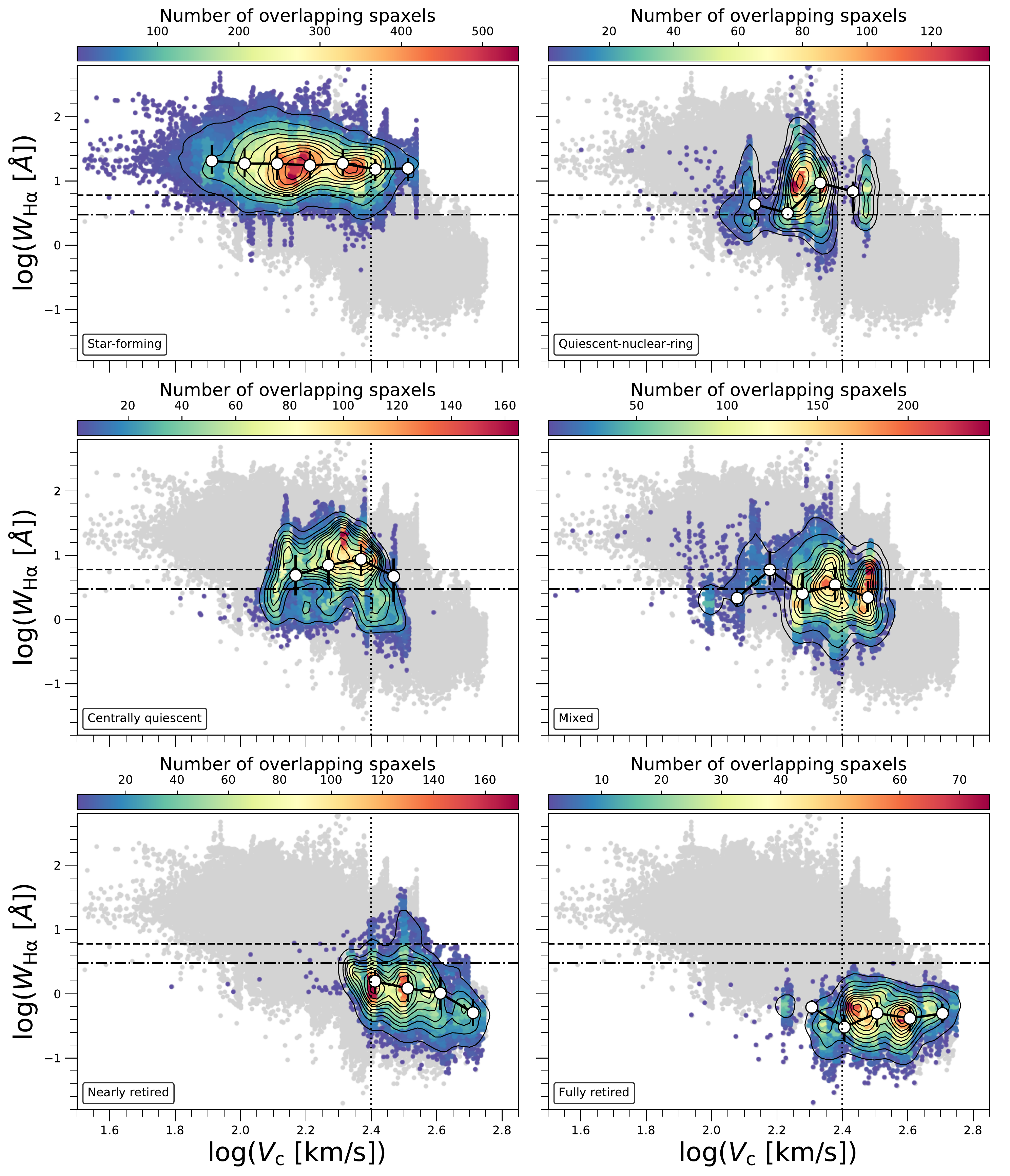}
\caption{Resolved $W_{\rm H\alpha}-V_{\rm c}$ scaling relation across quenching stages. Symbols and conventions follow Fig.~\ref{fig:Vc_Wha}. Grey dots indicates the location of all data points in the sample.} 
\label{fig:wha_vc_qs}
\end{figure*}

\begin{figure*}
\centering
\includegraphics[width=1\textwidth]{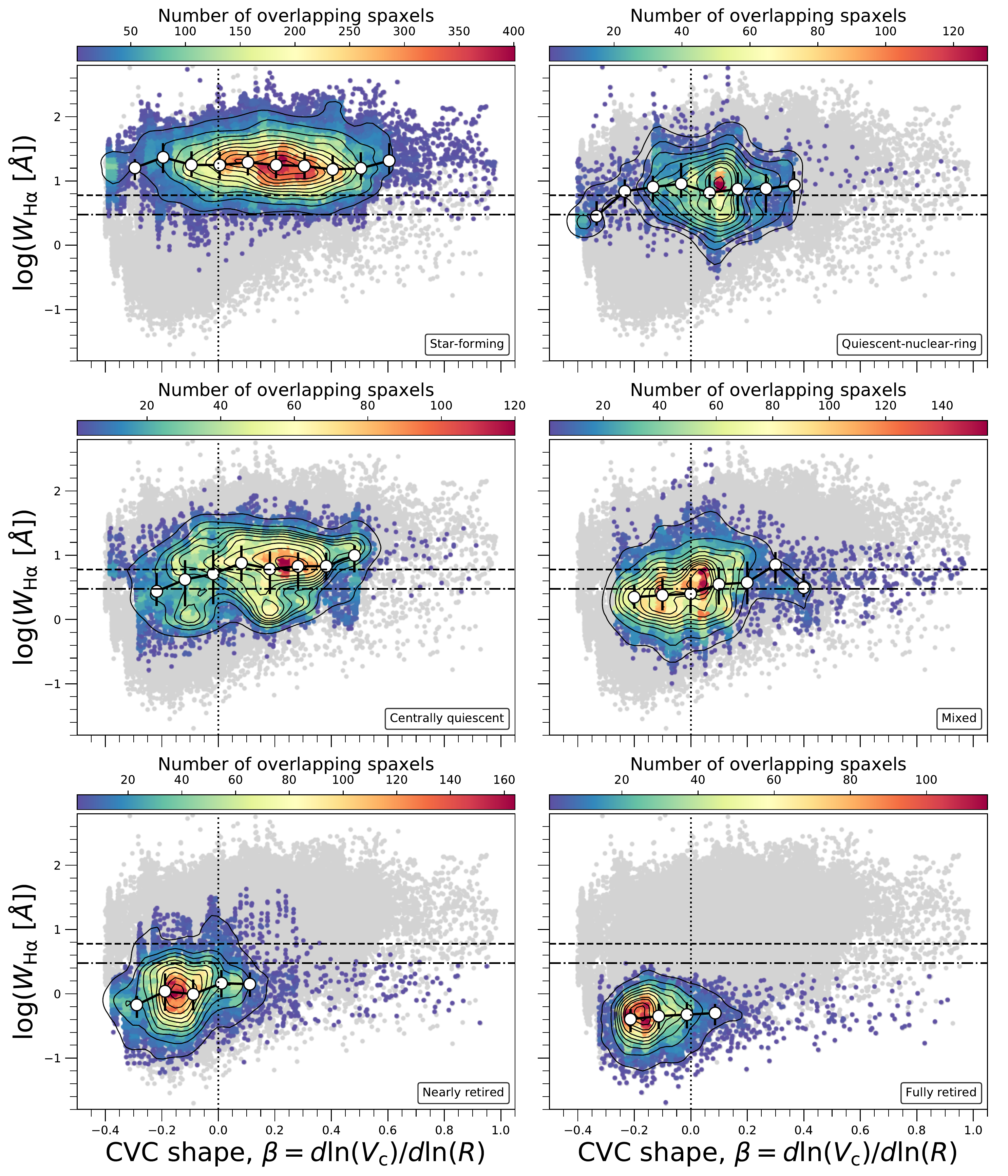}
\caption{Resolved $W_{\rm H\alpha}-\beta$ scaling relation across quenching stages. Symbols and conventions follow Fig.~\ref{fig:Beta_Wha}. Grey dots indicates the location of all data points in the sample.} 
\label{fig:wha_beta_qs}
\end{figure*}

Figures \ref{fig:wha_vc_qs} and \ref{fig:wha_beta_qs} present  the individual spaxel distribution of the resolved relationships $W_{\rm{H_{\alpha}}} - V_c$ and $W_{\rm{H_{\alpha}}} - \beta$ for each of the sixth quenching stages, respectively. We compared each of these distributions with the global distribution of the relationships, consisting of the spaxels of all quenching stages.  

\section{Stability of the $W_{\rm{H_{\alpha}}} - V_c$ and and $W_{\rm{H_{\alpha}}} - \beta$ relations}
\label{A:stability}
\begin{figure*}
\centering
\includegraphics[width=\textwidth]{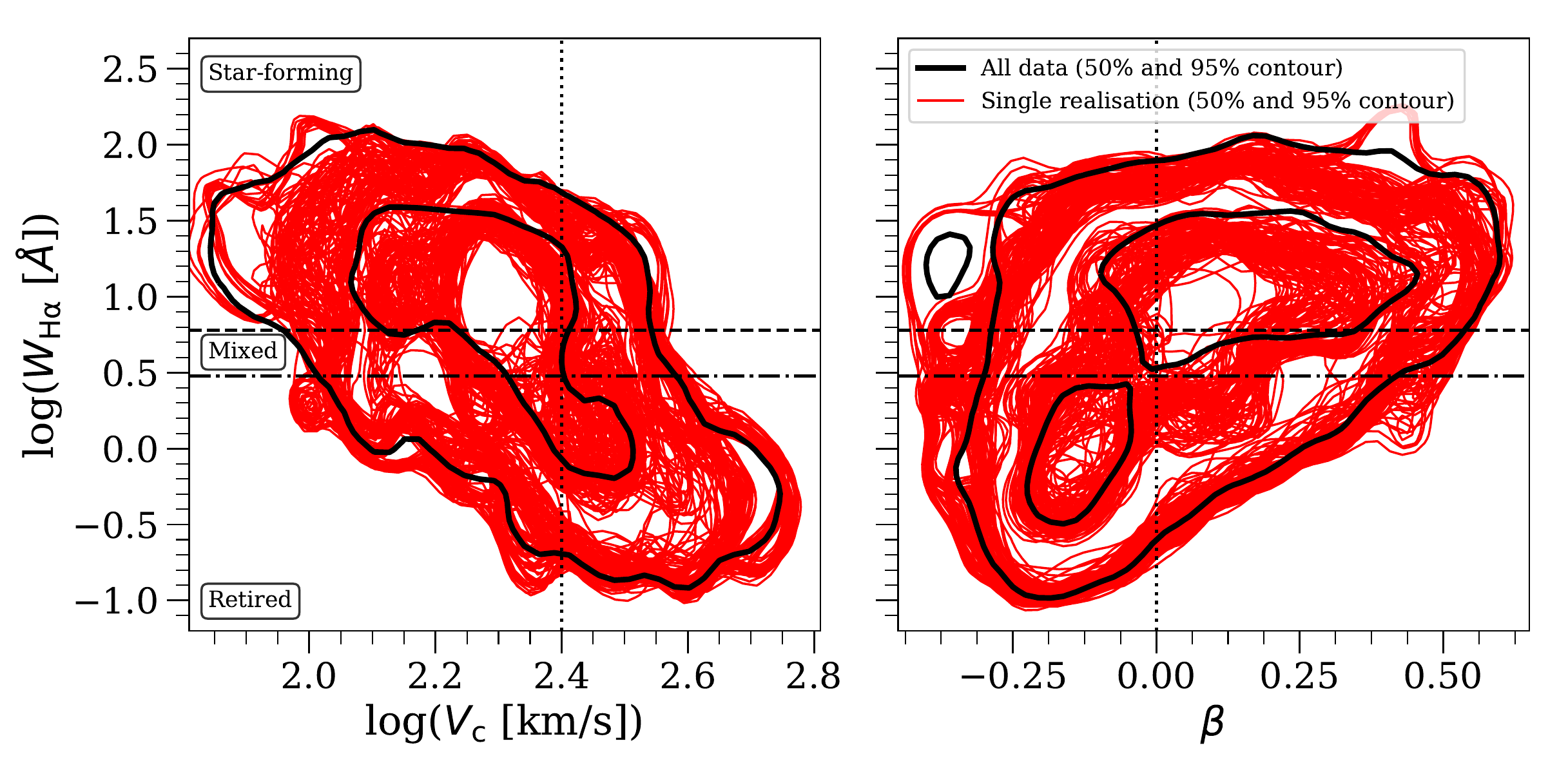}
\caption{Testing the stability of the resolved scaling relations ($W_{\rm H\alpha}-V_{\rm c}$, left; $W_{\rm H\alpha}-\beta$, right) through a cross-validation technique. The red lines indicate the spaxels of randomly selected 60 galaxies from the original data set (where 10 galaxies from each of the six quenching stages are chosen) for 100 single realisations. The inner and the outer contours represent the relationships constructed with 50\% and  95\% of the spaxel distribution, respectively, of the full sample of 215 galaxies (black contours) and each sub-sample of 60 galaxies (red contours). Both relationships overall show stability regarding the studied sample across quenching stage.}
\label{fig:cross_valid}
\end{figure*}

Both resolved relationships $W_{\rm{H_{\alpha}}} - V_c$ and $W_{\rm{H_{\alpha}}} - \beta$ show overall stability after verification with a cross-validation technique. We randomly select a sub-sample of 10 galaxies from each quenching stage group  (in order to keep the balance between star-forming and quenched systems) from the original data set, and reconstruct the two relationships via 100 realisations (i.e.  60 galaxies in total for each realisation) as shown in Fig. \ref{fig:cross_valid}. 

\begin{figure*}
\centering
\includegraphics[width=\textwidth]{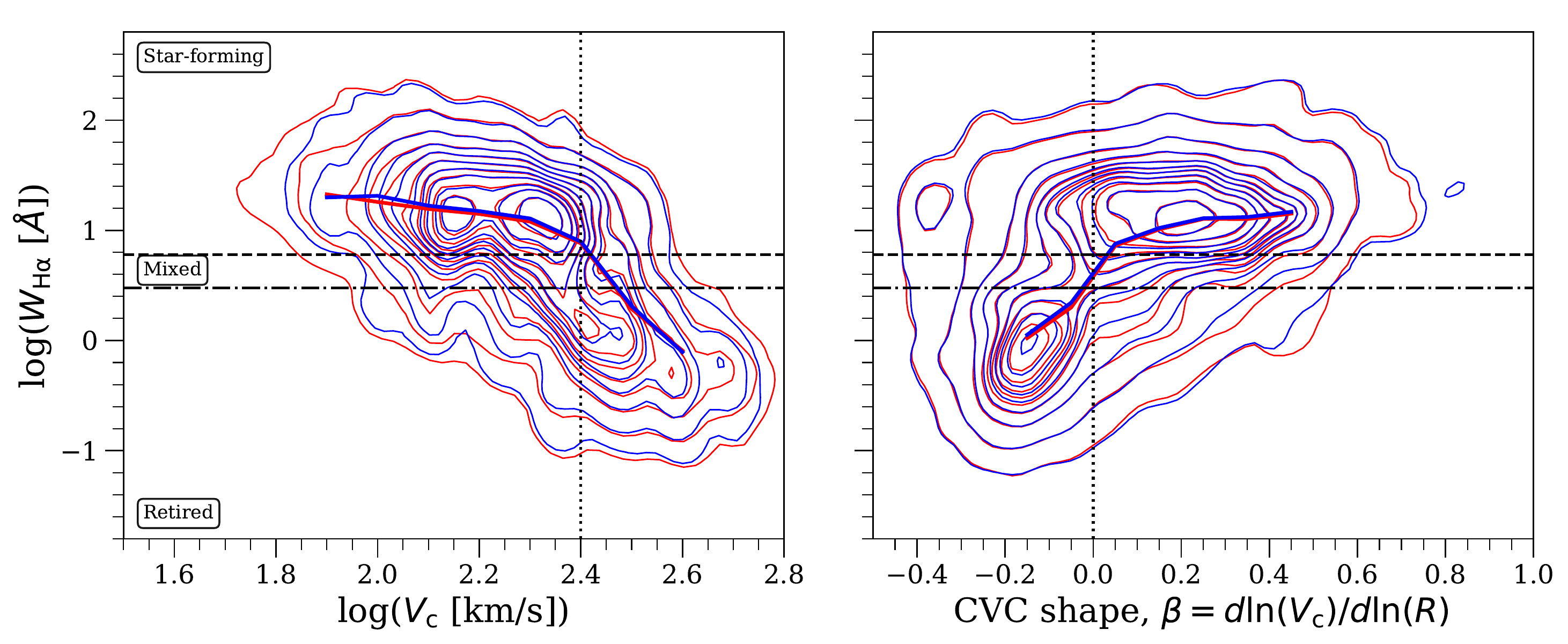}
\caption{Resolved scaling relations ($W_{\rm H\alpha}-V_{\rm c}$, left; $W_{\rm H\alpha}-\beta$, right) for all data in the sample ($R/R_{\rm e}>0$, red contours) and for data where $R/R_{\rm e}>0.5$ (blue contours). With the same colours, full lines indicate the running $W_{\rm H\alpha}$ related to the quantity on the $x-$axis for the given group.} 
\label{fig:scalrel_rad}
\end{figure*}

Furthermore, we test whether there is a significant contribution from the bulge region to a certain area in the $W_{\rm{H_{\alpha}}} - V_c$ and $W_{\rm{H_{\alpha}}} - \beta$ relations (especially in the quenching regions due to the expected act of the dynamical suppression mechanism). Interestingly,  the stability of the spaxel distribution of the relationships is preserved in both cases when the spaxels of the galaxies are taken into account with or without the central regions (i.e., below and above 0.5 R$_e$, respectively) as presented in Fig. \ref{fig:scalrel_rad}. This shows that the central spaxels of the galaxies are not necessary contributing  to any specific area in the global relationships (neither to a certain quenching areas).

\label{A:realisations}

\label{A:scalrel_rad}


\label{lastpage}
\end{document}